\documentclass[aps,twocolumn,prx]{revtex4-2}

\usepackage{xcolor,hyperref}
\hypersetup{
   colorlinks,
   linkcolor={blue!50!black},
   citecolor={blue!50!black},
   urlcolor={blue!80!black}
}

\usepackage{epsfig, amsmath, graphicx}
\usepackage{bm}
\usepackage{color}

\usepackage[latin1]{inputenc}
\newcommand{\beq}{\begin{equation}}
\newcommand{\eeq}{\end{equation}}
\newcommand{\bea}{\begin{eqnarray}}
\newcommand{\eea}{\end{eqnarray}}

\renewcommand{\=}{\!=\!}

\global\long\def\abs#1{\left|#1\right|}

\def\g-blue#1{\textcolor{g-blue}{#1}}

\begin{document}
\title{The dynamics of unsteady frictional slip pulses}
\author{Anna Pomyalov$^{1}$}
\author{Fabian Barras$^{2}$}
\author{Thibault Roch$^{3}$}
\author{Efim A.~Brener$^{4,5}$}
\email{e.brener@fz-juelich.de}
\author{Eran Bouchbinder$^{1}$}
\email{eran.bouchbinder@weizmann.ac.il}
\affiliation{$^{1}$Chemical and Biological Physics Department, Weizmann Institute of Science, Rehovot 7610001, Israel\\
$^{2}$The Njord Centre, Departments of Physics and Geosciences, University of Oslo, 0316 Oslo, Norway\\
$^{3}$Civil Engineering Institute, Materials Science and Engineering Institute, Ecole Polytechnique F\'ed\'erale de Lausanne, Station 18, CH-1015 Lausanne, Switzerland\\
$^{4}$Peter Gr\"unberg Institut, Forschungszentrum J\"ulich, D-52425 J\"ulich, Germany\\
$^{5}$Institute for Energy and Climate Research, Forschungszentrum J\"ulich, D-52425 J\"ulich, Germany}
\begin{abstract}
Self-healing slip pulses are major spatiotemporal failure modes of frictional systems, featuring a characteristic size $L(t)$ and a propagation velocity $c_{\rm p}(t)$ ($t$ is time). Here, we develop a theory of slip pulses in realistic rate-and-state dependent frictional systems. We show that slip pulses are intrinsically unsteady objects --- in agreement with previous findings --- yet their dynamical evolution is closely related to their unstable steady-state counterparts. In particular, we show that each point along the time-independent $L^{\mbox{\tiny{(0)}}}(\tau_{\rm d})\!-\!c^{\mbox{\tiny{(0)}}}_{\rm p}(\tau_{\rm d})$ line, obtained from a family of steady-state pulse solutions parameterized by the driving shear stress $\tau_{\rm d}$, is unstable. Nevertheless, and remarkably, the $c^{\mbox{\tiny{(0)}}}_{\rm p}[L^{\mbox{\tiny{(0)}}}]$ line is a dynamic attractor such that the unsteady dynamics of slip pulses (when they exist) --- whether growing ($\dot{L}(t)\!>\!0$) or decaying ($\dot{L}(t)\!<\!0$) --- reside on the steady-state line. The unsteady dynamics along the line are controlled by a single slow unstable mode. The slow dynamics of growing pulses, manifested by $\dot{L}(t)/c_{\rm p}(t)\!\ll\!1$, explain the existence of sustained pulses, i.e.~pulses that propagate many times their characteristic size without appreciably changing their properties. Our theoretical picture of unsteady frictional slip pulses is quantitatively supported by large-scale, dynamic boundary-integral method simulations.
\end{abstract}

\maketitle

Frictional systems, such as geological faults and the tectonic plates that form them, undergo interfacial failure through various spatiotemporal rupture modes that mediate rapid sliding. The main rupture modes are known to be expanding crack-like rupture and self-healing slip pulses~\cite{Ben-Zion2001,Scholz2002}. Self-healing slip pulses are solitonic structures that feature a characteristic size (compact support) $L(t)$ and a propagation velocity $c_{\rm p}(t)$, where $t$ is time~\cite{Freund1979,Heaton1990,Perrin1995,Beroza_Mikumo_1996,Beeler1996,Cochard1996,Andrews1997,Zheng1998,Nielsen2000,Nielsen2003,Brener2005,Shi2008,Rubin2009,Garagash2012,Gabriel2012,galetzka2015slip,Putelat2017,Michel2017,Brener2018,brantut2019stability,lambert2021propagation,ROCH2022104607,pomyalov2023self}. The existence of self-healing slip pulses --- having no counterparts in the tensile failure of materials --- crucially depends on the nonequilibrium nature of frictional interfaces, most notably on the ability of contact interfaces to recover their strength after undergoing sliding-induced weakening, i.e.~to self heal~\cite{Dieterich1978,Dieterich1994a,Beeler1994,Marone1998a,Berthoud1999,Baumberger2006Solid,Ben-David2010}.

Frictional slip pulses pose significant physical and mathematical challenges; even obtaining steady-state pulse solutions for realistic, laboratory-derived interfacial constitutive relations and studying their dynamic stability remain major open challenges. Previous work has shown that steady-state slip pulses are unstable objects~\cite{Perrin1995,Brener2018,brantut2019stability}. At the same time, observations show that slip pulses can be dynamically sustained, i.e.~to propagate over distances that are much larger than their characteristic size without appreciably changing their shape and propagation velocity~\cite{Gabriel2012,Brener2018}. This combined evidence may appear intrinsically conflicting, reflecting the lack of current understanding of the dynamics of unsteady frictional slip pulses. This problem is of prime importance in fields such as nonequilibrium and nonlinear physics, materials physics, tribology and geophysics.

Here, we develop a theoretical picture of unsteady frictional slip pulses, and reveal a surprising intrinsic relation between them and their unstable steady-state counterparts. Generic rate-and-state dependent frictional systems admit a family of steady-state slip pulse solutions parameterized by the driving shear stress $\tau_{\rm d}$, featuring a size-velocity $L^{\mbox{\tiny{(0)}}}(\tau_{\rm d})\!-\!c^{\mbox{\tiny{(0)}}}_{\rm p}(\tau_{\rm d})$ relation (the superscript $^{\mbox{\tiny{(0)}}}$ refers to time-independent steady-state pulse solutions)~\cite{pomyalov2023self}. We show that each point along the $c^{\mbox{\tiny{(0)}}}_{\rm p}[L^{\mbox{\tiny{(0)}}}]$ line is unstable. That is, slightly perturbed steady-state pulses either grow in time ($\dot{L}(t)\!>\!0$) or decay ($\dot{L}(t)\!<\!0$) such that steady-state pulses correspond to a saddle configuration. Yet, the resulting unsteady slip pulses (when they are excited, and not crack-like rupture) do not follow arbitrary dynamics in the $L(t)\!-\!c_{\rm p}(t)$ plane as a function of time, but are rather attracted toward and closely adhere to the $c^{\mbox{\tiny{(0)}}}_{\rm p}[L^{\mbox{\tiny{(0)}}}]$ line. Consequently, one is able to understand the dynamics of unsteady slip pulses.

Furthermore, the unsteady dynamics along the $c^{\mbox{\tiny{(0)}}}_{\rm p}[L^{\mbox{\tiny{(0)}}}]$ line are controlled by a single slow unstable mode. In particular, growing pulses feature $\dot{L}(t)/c_{\rm p}(t)\!\ll\!1$ and are nothing but sustained pulses that propagate many times their characteristic size without appreciably changing their properties. The $c^{\mbox{\tiny{(0)}}}_{\rm p}[L^{\mbox{\tiny{(0)}}}]$ line is a dynamic attractor for unsteady slip pulses independently of their initial distance from the line, i.e.~the theory applies to general initial conditions that lead to unsteady slip pulses. Our theoretical picture of unsteady frictional slip pulses is quantitatively supported by large-scale, dynamic boundary-integral method simulations for various initial conditions.

{\em The dynamic stability of steady-state pulse solutions}.--- The frictional strength $\tau$ of contact interfaces depends on the time rate of change of the tangential displacement discontinuity across the interface, i.e.~on the slip velocity $v(x,t)$ (here $x$ is the coordinate along the interface), and on the structural state of the interface. As macroscopic contact interfaces are never smooth on all scales~\cite{Dieterich1994a,Marone1998a,Baumberger2006Solid,bowden2001friction}, a central structural state field is the real contact area, composed of an ensemble of contact asperities, which is typically orders of magnitude smaller than the nominal contact area~\cite{Dieterich1994a,bowden2001friction}. Moreover, this scale separation between the real and nominal contact area implies that contact asperities experience intense loads and typically age with time. The aging time $\phi(x,t)$ is commonly used to quantify the nonequilibrium structural state of the interface~\cite{Marone1998a,Baumberger2006Solid}.
\begin{figure}[ht!]
\centering
\includegraphics[width=1\columnwidth]{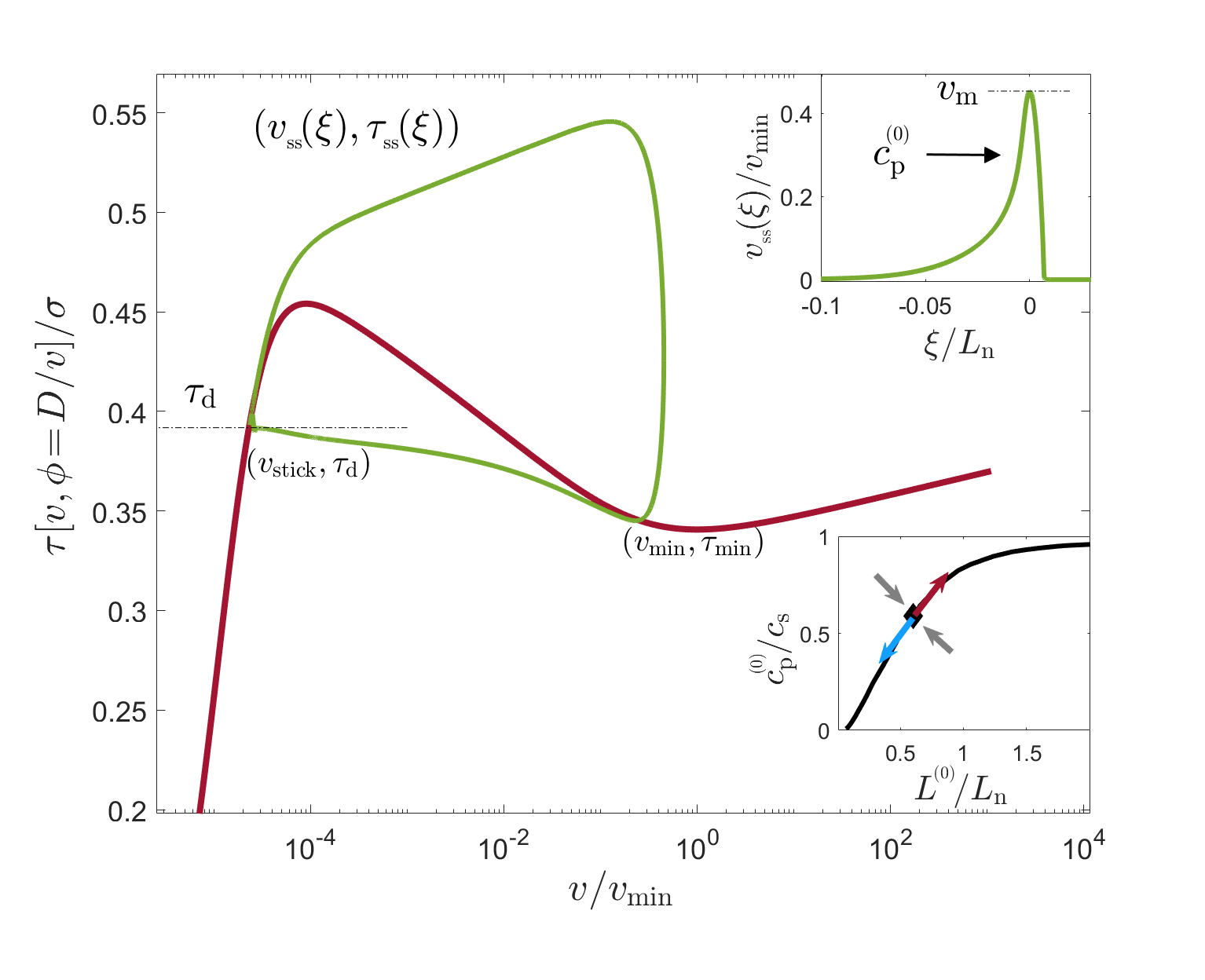}
\caption{(main) The steady frictional resistance $\tau[v,\phi\!=\!D/v]$ (solid brown line, in units of the normal stress $\sigma$) vs.~slip velocity $v$ (in units of the $v_{\rm min}$, the velocity at the local minimum). A steady-state slip pulse corresponds to a closed (homoclinic) orbit $(v_{_{\rm ss}}\!(\xi),\tau_{_{\rm ss}}\!(\xi))$ (solid green line) parameterized by the co-moving space-time coordinate $\xi\!=\!x-c^{\mbox{\tiny{(0)}}}_{\rm p}t$, where $c^{\mbox{\tiny{(0)}}}_{\rm p}$ is the steady pulse propagation velocity. The closed orbit starts and ends at $(v_{\rm stick},\tau_{\rm d})$, where $\tau_{\rm d}$ is the driving shear stress and $v_{\rm stick}$ is extremely small. The existence of a minimum $(v_{\rm min},\tau_{\rm min})$ of the friction curve, whose values  are used to normalize various quantities, does not have a qualitative effect on the results. Yet, it might affect quantitative aspects of unsteady slip pulse propagation. (upper inset) $v_{_{\rm ss}}\!(\xi)$, which features a maximal value $v_{\rm m}$ and a characteristic size $L^{\mbox{\tiny{(0)}}}$ (not marked, see~\cite{SM}), is plotted ($\xi$ is expressed in units of $L_{\rm n}$, a natural normalization length~\cite{SM}). (lower inset) The size-velocity relation $c^{\mbox{\tiny{(0)}}}_{\rm p}[L^{\mbox{\tiny{(0)}}}]$ (in units of the shear wave-speed $c_{\rm s}$) for a family of steady-state pulse solutions parameterized by $\tau_{\rm d}$ is plotted (solid line). A solution for a given $\tau_{\rm d}$ is marked by the diamond, and its expected stability properties are represented by the arrows (see text for discussion).}
\label{fig:fig1}
\end{figure}

Consequently, the frictional strength is a functional $\tau[v(x,t),\phi(x,t)]$, whose form is extracted from experimental measurements and theoretical considerations~\cite{Marone1998a,Baumberger2006Solid,Bar-Sinai2014}, see Fig.~\ref{fig:fig1}. Finally, contact aging in the absence of sliding ($v\=0$) and sliding-induced ($v\!>\!0$) rejuvenation (anti-aging) are described as $\partial_t\phi(x,t)\=1\!-\!v(x,t)\,\phi(x,t)/D$, where the transition between the two rheological behaviors is controlled by a characteristic slip displacement $D$~\cite{Dieterich1979a,Ruina1983,Nakatani2001,Marone1998a,Baumberger2006Solid}. The frictional strength $\tau[v(x,t),\phi(x,t)]$ balances the shear stress emerging from the deformation of the bodies forming the contact interface, at any point $x$ and time $t$. The deformable bodies extend in the $y$ direction (the interface is located at $y\=0$), and are assumed to be translationally invariant along the third direction $z$, making the problem considered here two-dimensional. When the bodies forming the contact interface are linear elastic and of infinite extent (or when the dynamics of interest are such that waves reflected from finite boundaries do not have time to affect the interface), the interfacial shear stress can be expressed in terms of the elastodynamic Green's function.

Considering then a frictional system driven anti-symmetrically by a shear stress $\tau_{\rm d}$ applied far from the interface, interfacial stress balance takes the form~\cite{das1980numerical}
\begin{equation}
\tau[v(x,t),\phi(x,t)] = \tau_{\rm d} -\frac{\mu}{2c_s}v(x,t) + s(x,t) \ .
\label{eq:BIM}
\end{equation}
Here, $\mu$ and $c_s$ are the shear modulus and wave-speed of the bodies, respectively, and $s(x,t)$ is a spatiotemporal convolutional integral that accounts for the long-range interaction of different parts of the interface, mediated by bulk deformation. In general, $s(x,t)$ does not admit an explicit real-space representation, and its specific form is different for anti-plane shear symmetry (i.e.~slip displacement in the $z$ direction) and in-plane shear symmetry (i.e.~slip displacement in the $x$ direction)~\cite{Breitenfeld1998}. Together with the $\partial_t\phi(x,t)$ equation stated above and an explicit expression for $\tau[v(x,t),\phi(x,t)]$~\cite{SM} (see Fig.~\ref{fig:fig1}), the coupled bulk-interface frictional dynamics are described by the specified set of nonlinear integro-differential equations.

Self-healing slip pulses correspond to solutions to the set of nonlinear integro-differential equations, which propagate at an instantaneous velocity $c_{\rm p}(t)$ and feature a finite length $L(t)$. The compact support $L(t)$, over which slip velocities are large, separates two spatially homogeneous, nearly quiescent, identical interfacial states. The latter are characterized by a vanishingly small slip velocity $v_{\rm stick}\!\to\!0$ (`stick' denotes the fact that the state is nearly quiescent). It is obtained by solving $\tau_{\rm d}\=\tau[v_{\rm stick},D/v_{\rm stick}]$ for a given driving shear stress $\tau_{\rm d}$, where $\phi\=D/v$ corresponds to $\partial_t\phi\=0$. In the language of dynamical systems, self-healing slip pulses correspond to closed (homoclinic) orbits in the $v\!-\!\phi$ plane, which both starts and ends at $(v_{\rm stick},D/v_{\rm stick})$. Alternatively, they correspond to closed (homoclinic) orbits in the $v\!-\!\tau$ plane, which both start and end at $(v_{\rm stick},\tau_{\rm d})$, see Fig.~\ref{fig:fig1}.

Steady-state slip pulses correspond to solutions that also satisfy $c_{\rm p}(t)\=c^{\mbox{\tiny{(0)}}}_{\rm p}$, i.e.~which propagate steadily at a time-independent velocity $c^{\mbox{\tiny{(0)}}}_{\rm p}$. Under such conditions, all fields depend on the co-moving space-time coordinate $\xi\!\equiv\!x-c^{\mbox{\tiny{(0)}}}_{\rm p}t$ and one has $\partial_t\=-c^{\mbox{\tiny{(0)}}}_{\rm p}\partial_\xi$ (applied to the $\partial_t\phi(x,t)$ equation). Equation~\eqref{eq:BIM} then takes the form~\cite{Weertman1980}
\begin{equation}
\tau[v_{_{\rm ss}}\!(\xi),\phi_{_{\rm ss}}\!(\xi)] = \tau_{\rm d} - \frac{\mu\,{\cal F}(c^{\mbox{\tiny{(0)}}}_{\rm p}/c_s)}{c^{\mbox{\tiny{(0)}}}_{\rm p}}\!\int_{-\infty}^{\infty}\frac{v_{_{\rm ss}}\!(z)}{z-\xi}dz \ ,
\label{eq:ss}
\end{equation}
where the integral (which should be understood in the Cauchy principal value sense) is the steady-state limit of the two last terms on the right-hand-side of ~\eqref{eq:BIM} and ${\cal F}(c^{\mbox{\tiny{(0)}}}_{\rm p}/c_s)$ is a known function (different for anti-plane and in-plane shear symmetries)~\cite{Weertman1980}.

A complete family of steady-state slip pulse solutions, i.e.~$v_{_{\rm ss}}\!(\xi)$, $\phi_{_{\rm ss}}\!(\xi)$ and $c^{\mbox{\tiny{(0)}}}_{\rm p}$ parameterized by $\tau_{\rm d}$, has been very recently obtained~\cite{pomyalov2023self}, see an example in the upper inset in Fig.~\ref{fig:fig1}. The properties of the solutions and their theoretical understanding have been discussed in~\cite{pomyalov2023self}. Most relevant for our purposes here is that the family of steady-state slip pulse solutions features a size-velocity $L^{\mbox{\tiny{(0)}}}(\tau_{\rm d})\!-\!c^{\mbox{\tiny{(0)}}}_{\rm p}(\tau_{\rm d})$ relation that is parameterized by $\tau_{\rm d}$, where $L^{\mbox{\tiny{(0)}}}$ is the characteristic size of each pulse~\cite{SM}. This relation is monotonically increasing and is plotted in the lower inset in Fig.~\ref{fig:fig1} for anti-plane shear symmetry (and a representative set of frictional constitutive parameters~\cite{SM}).

In principle, the physical relevance of the obtained steady-state slip pulse solutions should be judged based on their dynamic stability. The above formulated problem, which corresponds to bodies of infinite height $H$ (i.e.~$H\!\to\!\infty$), has been previously solved in the limit of small $H$ (where long-range elastodynamic interactions become local)~\cite{Brener2018}. It has been shown that steady-state slip pulses in this limit are intrinsically unstable, i.e.~that $L^{\mbox{\tiny{(0)}}}(\tau_{\rm d})+\delta{L}$ perturbations (with $\delta{L}\!>\!0$) lead to growing pulses, while $L^{\mbox{\tiny{(0)}}}(\tau_{\rm d})-\delta{L}$ perturbations lead to decaying pulses. If the two problems are smoothly connected by continuously increasing $H$, we expect slip pulses in the $H\!\to\!\infty$ limit to be intrinsically unstable as well --- in agreement with existing literature~\cite{Perrin1995,brantut2019stability} ---, i.e.~to correspond to a saddle configuration for each $\tau_{\rm d}$. This expectation is tested next in the present context.
\begin{figure}[ht!]
\centering
\includegraphics[width=\columnwidth]{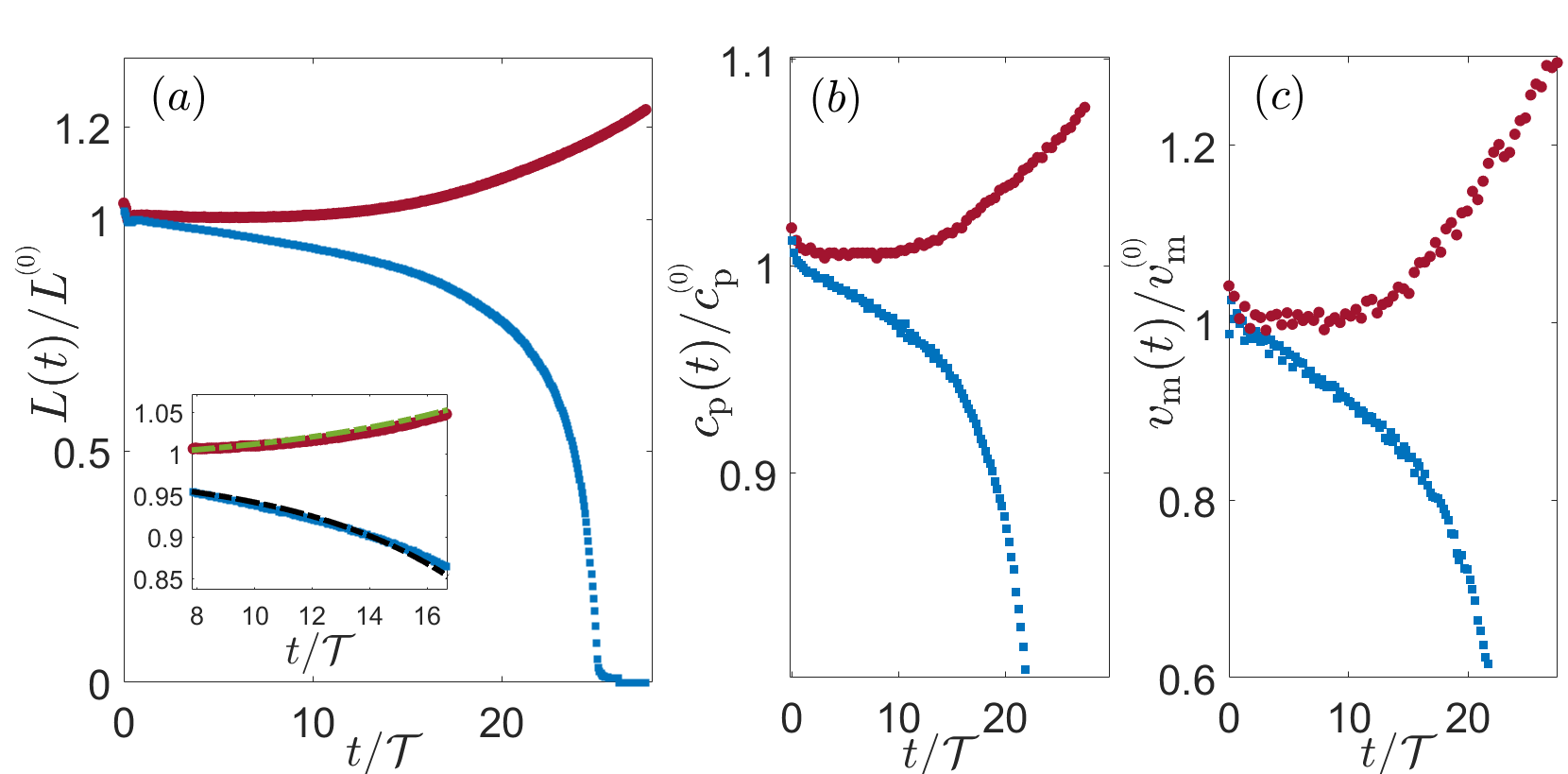}
\caption{Dynamic stability analysis of a steady-state pulse with $\tau_{\rm d}\!=\!1.03\tau_{\rm min}\!\approx\!0.351\sigma$ (featuring $c^{\mbox{\tiny{(0)}}}_{\rm p}\!=\!0.82c_{\rm s}$) using the $\phi(x,t)$ field $\epsilon$-perturbation scheme described in the text, with $\epsilon\!=\!-0.15$ (brown data) and $\epsilon\!=\!-0.05$ (blue data). (a) $L(t)/L^{\mbox{\tiny{(0)}}}$ vs.~$t/{\cal T}$. (b) $c_{\rm p}(t)/c^{\mbox{\tiny{(0)}}}_{\rm p}$ vs.~$t/{\cal T}$. (c) $v_{\rm m}(t)/v^{\mbox{\tiny{(0)}}}_{\rm m}$ vs.~$t/{\cal T}$, with ${\cal T}\!=\!L^{\mbox{\tiny{(0)}}}/c^{\mbox{\tiny{(0)}}}_{\rm p}$. For all 3 quantities, $\epsilon\!=\!-0.15$ (brown data) leads to growth and $\epsilon\!=\!-0.05$ (blue data) to decay, demonstrating the unstable nature of the steady-state pulse. The timescale of instability is similar in all quantities and is set by ${\cal T}$. In the inset of panel (a), we test the predicted unstable linearized dynamics in ~\eqref{eq:L_dynamics_linear}. Since $\epsilon$ is not exactly centered around $\epsilon_{\rm c}\!=\!0$, we consider the solution of ~\eqref{eq:L_dynamics_linear} in the form $L_{\pm}(t)\!=\!L^{\mbox{\tiny{(0)}}}\!\pm\!\Delta{L}\exp[a_{\pm}(t-t_1)/{\cal T}]$ for $t\!>\!t_1$, with $t_1\!\simeq\!8{\cal T}$, where $\pm$ correspond to the upper/lower branches, respectively. The results with $\Delta{L}/L^{\mbox{\tiny{(0)}}}\!\simeq\!0.01$ (quantifying the deviation from steady state at $t_1$) are superimposed, for the upper branch with $a_{+}\!=\!0.13$ (green dashed-dotted line) and for the lower one with $a_{-}\!=\!0.17$ (black dashed-dotted line), supporting the prediction in ~\eqref{eq:L_dynamics_linear}.}
\label{fig:fig2}
\end{figure}

To test this expectation, we use the steady-state pulse solutions --- $v_{_{\rm ss}}\!(\xi)$, $\phi_{_{\rm ss}}\!(\xi)$ and $c^{\mbox{\tiny{(0)}}}_{\rm p}$ for a given $\tau_{\rm d}$ --- as initial conditions for the entire history $-\infty\!<\!t\!<\!0$ in ~\eqref{eq:BIM} and for the $\partial_t\phi(x,t)$ equation, slightly perturb them in both directions ($\pm\delta{L}$) at $t\=0$ and follow the subsequent dynamics using large-scale, dynamic boundary-integral method simulations~\cite{das1980numerical,Geubelle1995}, using the open-source library cRacklet~\cite{roch2022cracklet}, see details in~\cite{SM}. The perturbation at $t\=0$ takes the form $\phi(x,t\=0)\=(1+\epsilon)\phi(x,t\=0^-)$. That is, we perturb the state field $\phi(x,t)$, where $\epsilon\!<\!0$ corresponds to $+\delta{L}$ perturbations and $\epsilon\!>\!0$ to $-\delta{L}$ ones. The results are presented in Fig.~\ref{fig:fig2}, where $L(t\!>\!0)$, $c_{\rm p}(t\!>\!0)$ and the maximal slip velocity $v_{\rm m}(t\!>\!0)$ (see upper inset in Fig.~\ref{fig:fig1}) are plotted for $\epsilon\=-0.15$ and $\epsilon\=-0.05$. It is observed that for $\epsilon\=-0.15$ the pulse grows, while for $\epsilon\=-0.05$ it decays, where transition from decay to growth takes place at $\epsilon_{\rm c}$ in between. While theoretically we expect the transition to occur exactly at $\epsilon_{\rm c}\=0$, the numerical transition point is shifted by a few percent due to finite size effects and computational constraints (which limit the history duration, see~\cite{SM}).

The results in Fig.~\ref{fig:fig2} show that steady-state pulses are intrinsically unstable and correspond to a saddle configuration, separating growing from decaying pulses. The characteristic timescale of variation of all 3 quantities ($L(t)$, $c_{\rm p}(t)$ and $v_{\rm m}(t)$), to be denoted by ${\cal T}$, appears to be similar. Focusing, for example, on the time evolution of $L(t)$ near $L^{\mbox{\tiny{(0)}}}(\tau_{\rm d})$, we expect it to follow the linearized dynamics
\begin{equation}
\dot{L}(t)\!=\!\frac{L(t)-L^{\mbox{\tiny{(0)}}}(\tau_{\rm d})}{{\cal T}(\tau_{\rm d})} \,\,\quad\hbox{for}\quad\,\, \frac{L(t)\!-\!L^{\mbox{\tiny{(0)}}}(\tau_{\rm d})}{L^{\mbox{\tiny{(0)}}}(\tau_{\rm d})}\!\ll\!1\ .
\label{eq:L_dynamics_linear}
\end{equation}
Moreover, a dimensional estimate suggests that ${\cal T}(\tau_{\rm d})\!\sim\!L^{\mbox{\tiny{(0)}}}(\tau_{\rm d})/c^{\mbox{\tiny{(0)}}}_{\rm p}(\tau_{\rm d})$. These predictions are verified in the inset of Fig.~\ref{fig:fig2}a. Taken together, our results indicate that slip pulses are unsteady objects and that their dynamics near steady state is characterized by a single unstable mode, featuring an instability rate $\sim\!1/{\cal T}(\tau_{\rm d})$.

{\em Theory of unsteady frictional slip pulses}.--- The unstable nature of steady-state slip pulses in large systems, established above, may suggest that steady-state slip pulse solutions are largely irrelevant for understanding the unsteady dynamics of slip pulses in realistic frictional systems. That is, the results show that slip pulses (under a fixed $\tau_{\rm d}$) that are initially close to $(L^{\mbox{\tiny{(0)}}}(\tau_{\rm d}),c^{\mbox{\tiny{(0)}}}_{\rm p}(\tau_{\rm d}))$ are dynamically driven away from this point. Moreover, the obtained results appear to indicate that pulses that are unrelated to $(L^{\mbox{\tiny{(0)}}}(\tau_{\rm d}),c^{\mbox{\tiny{(0)}}}_{\rm p}(\tau_{\rm d}))$ to begin with, will reveal no special relation to the $c^{\mbox{\tiny{(0)}}}_{\rm p}[L^{\mbox{\tiny{(0)}}}]$ line, corresponding to unstable steady-state slip pulse solutions. If true, then a theory of unsteady frictional slip pulses should be developed independently of steady-state slip pulse solutions (which is a formidable task). We show that this is in fact not the case. That is, while all of the above statements regarding the unstable steady-state point $(L^{\mbox{\tiny{(0)}}}(\tau_{\rm d}),c^{\mbox{\tiny{(0)}}}_{\rm p}(\tau_{\rm d}))$ for a given $\tau_{\rm d}$ are valid, the entire $c^{\mbox{\tiny{(0)}}}_{\rm p}[L^{\mbox{\tiny{(0)}}}]$ line --- parameterized by $\tau_{\rm d}$ --- is of fundamental importance for unsteady frictional slip pulses.

To understand this, let us consider pulse dynamics in the $L(t)\!-\!c_{\rm p}(t)$ plane, where the steady-state $c^{\mbox{\tiny{(0)}}}_{\rm p}[L^{\mbox{\tiny{(0)}}}]$ line is defined (see lower inset of Fig.~\ref{fig:fig1}). That is, while slip pulses are characterized by the fields $v(x,t)$ and $\phi(x,t)$, i.e.~by infinitely many degrees of freedom, we are looking for a reduced-dimensionality description based on a small number of coarse-grained discrete variables, in particular two such variables. The specific choice of $L(t)$ and $c_{\rm p}(t)$ is physically motivated, as the pulse size and propagation velocity are of fundamental importance, yet it is by no means unique; in fact, we show that $L(t)$ and $v_{\rm m}(t)$ could have been used as well~\cite{SM}. The two-dimensional dynamical system for $\dot{L}(t)$ and $\dot{c}_{\rm p}(t)$ (under a given $\tau_{\rm d}$) is unknown and is in general nonlinear arbitrarily far from the point $(L^{\mbox{\tiny{(0)}}}(\tau_{\rm d}),c^{\mbox{\tiny{(0)}}}_{\rm p}(\tau_{\rm d}))$.

Yet, the short-time dynamics near $(L^{\mbox{\tiny{(0)}}}(\tau_{\rm d}),c^{\mbox{\tiny{(0)}}}_{\rm p}(\tau_{\rm d}))$ follow linearized equations as in ~\eqref{eq:L_dynamics_linear} along its unstable direction. As the dynamical system is two-dimensional and there is one unstable mode (formally corresponding to a positive eigenvalue), the other mode is stable (corresponding to a negative eigenvalue). It is natural to expect the direction of the unstable mode in the $L(t)\!-\!c_{\rm p}(t)$ plane to be along the local tangent to $c^{\mbox{\tiny{(0)}}}_{\rm p}[L^{\mbox{\tiny{(0)}}}]$ line at $(L^{\mbox{\tiny{(0)}}}(\tau_{\rm d}),c^{\mbox{\tiny{(0)}}}_{\rm p}(\tau_{\rm d}))$ and the stable direction to be perpendicular to it, as is illustrated by the outgoing and ingoing arrows in the lower inset in Fig.~\ref{fig:fig1}. Plotting the results of Fig.~\ref{fig:fig2}a-b in the $L(t)\!-\!c_{\rm p}(t)$ plane (see~\cite{SM}) indeed show that the unstable short-time dynamics near $(L^{\mbox{\tiny{(0)}}}(\tau_{\rm d}),c^{\mbox{\tiny{(0)}}}_{\rm p}(\tau_{\rm d}))$ evolve along the local tangent to the $c^{\mbox{\tiny{(0)}}}_{\rm p}[L^{\mbox{\tiny{(0)}}}]$ line at this point.

Since the above results apply to any point along the $c^{\mbox{\tiny{(0)}}}_{\rm p}[L^{\mbox{\tiny{(0)}}}]$ line, a rather remarkable picture emerges. Invoking the continuity of dynamical trajectories in $L(t)\!-\!c_{\rm p}(t)$ plane and assuming $c^{\mbox{\tiny{(0)}}}_{\rm p}[L^{\mbox{\tiny{(0)}}}]$ to be the only steady-state line, we expect all dynamical trajectories that are away from the $c^{\mbox{\tiny{(0)}}}_{\rm p}[L^{\mbox{\tiny{(0)}}}]$ line to be attracted to it and once they hit it, say at time $t_0$, to evolve along the $c^{\mbox{\tiny{(0)}}}_{\rm p}[L^{\mbox{\tiny{(0)}}}]$ line, whether corresponding to decaying or growing pulses. The latter implies that if one treats $L(t)$ as the ``independent variable'', we expect $c_{\rm p}(t)$ to be determined according to
\begin{equation}
\label{eq:solution}
c_{\rm p}(t) \simeq c^{\mbox{\tiny{(0)}}}_{\rm p}[L(t)] \qquad\hbox{for}\qquad t>t_0\ ,
\end{equation}
where the right-hand-side of ~\eqref{eq:solution} is the steady-state relation $c^{\mbox{\tiny{(0)}}}_{\rm p}[L^{\mbox{\tiny{(0)}}}]$ and its argument is the dynamic $L(t)$. Moreover, if $L(t_0)\!>\!L^{\mbox{\tiny{(0)}}}(\tau_{\rm d})$, then the pulse will grow ($\dot{L}(t\!>\!t_0)\!>\!0$), and if $L(t_0)\!<\!L^{\mbox{\tiny{(0)}}}(\tau_{\rm d})$, it will decay ($\dot{L}(t\!>\!t_0)\!<\!0$).

The emerging physical picture of unsteady slip pulses, to be quantitatively tested below, reveals that while each point $(L^{\mbox{\tiny{(0)}}}\!(\tau_{\rm d}),c^{\mbox{\tiny{(0)}}}_{\rm p}\!(\tau_{\rm d}))$ along the steady-state $c^{\mbox{\tiny{(0)}}}_{\rm p}[L^{\mbox{\tiny{(0)}}}]$ line is unstable in itself, i.e.~the dynamics are repelled from it (as indicated by the arrows in the lower inset in Fig.~\ref{fig:fig1}), the entire steady-state $c^{\mbox{\tiny{(0)}}}_{\rm p}[L^{\mbox{\tiny{(0)}}}]$ line is a dynamic attractor for unsteady slip pulses, whether decaying or growing.

{\em Sustained slip pulses}.--- Growing pulses are of particular importance as they are accompanied by large slip and radiated energy, and hence would play major roles in the failure of frictional interfaces. Indeed, many experimental and numerical works demonstrate the existence of such unsteady slip pulses~\cite{Freund1979,Heaton1990,Perrin1995,Beroza_Mikumo_1996,Beeler1996,Cochard1996,Andrews1997,Zheng1998,Nielsen2000,Nielsen2003,Brener2005,Shi2008,Rubin2009,Garagash2012,Gabriel2012,galetzka2015slip,Putelat2017,Michel2017,Brener2018,brantut2019stability,lambert2021propagation,ROCH2022104607}. It is observed that while these pulses are indeed out of steady state (as our theory predicts), they are nevertheless ``sustained'' in the sense that they propagate many times their characteristic size without appreciably changing their properties. Our theory of unsteady pulses may offer a natural explanation for sustained slip pulses. Clearly, we focus on pulses that correspond to $L(t_0)\!>\!L^{\mbox{\tiny{(0)}}}(\tau_{\rm d})$, leading to growth, i.e.~$\dot{L}(t)\!>\!0$ for $t\!>\!t_0$. If these are indeed sustained pulses, then the unstable mode (discussed above) that controls their growth should be a slow mode.

To quantify the possible slowness of the growing unstable mode, note that the time it takes a growing pulse to propagate a distance comparable to its own size is $\Delta{t}\!\simeq\!L(t)/c_{\rm p}(t)$. The latter is valid if the change in size over this time $\Delta{L}(t)\!\simeq\!\Delta{t}\,\dot{L}(t)$ is much smaller than $L(t)$, which is a dimensionless measure of the slowness of the unstable mode. The expectation that $\Delta{L}(t)\!\ll\!L(t)$ amounts to
\begin{equation}
\label{eq:slow}
    \dot{L}(t)/c_{\rm p}(t)\!\ll\!1  \,\,\qquad\hbox{for}\qquad\,\, \dot{L}(t)\!>\!0
\end{equation}
for $t\!>\!t_0$, which will be also quantitatively tested below.

{\em Boundary-integral method simulational support}.--- Our next goal is to quantitatively test the theoretical predictions discussed above in large-scale, dynamic boundary-integral method simulations~\cite{roch2022cracklet,SM}. This is achieved in two steps. First, we aim at testing the predictions that: (i) pulses that are initially close to the steady-state $c^{\mbox{\tiny{(0)}}}_{\rm p}[L^{\mbox{\tiny{(0)}}}]$ line remain close to it at later times and (ii) that the position of the initial point relative to $(L^{\mbox{\tiny{(0)}}}\!(\tau_{\rm d}),c^{\mbox{\tiny{(0)}}}_{\rm p}\!(\tau_{\rm d}))$ determines whether a pulse grows or decays along the line.
\begin{figure}[ht!]
\centering
\includegraphics[width=1\columnwidth]{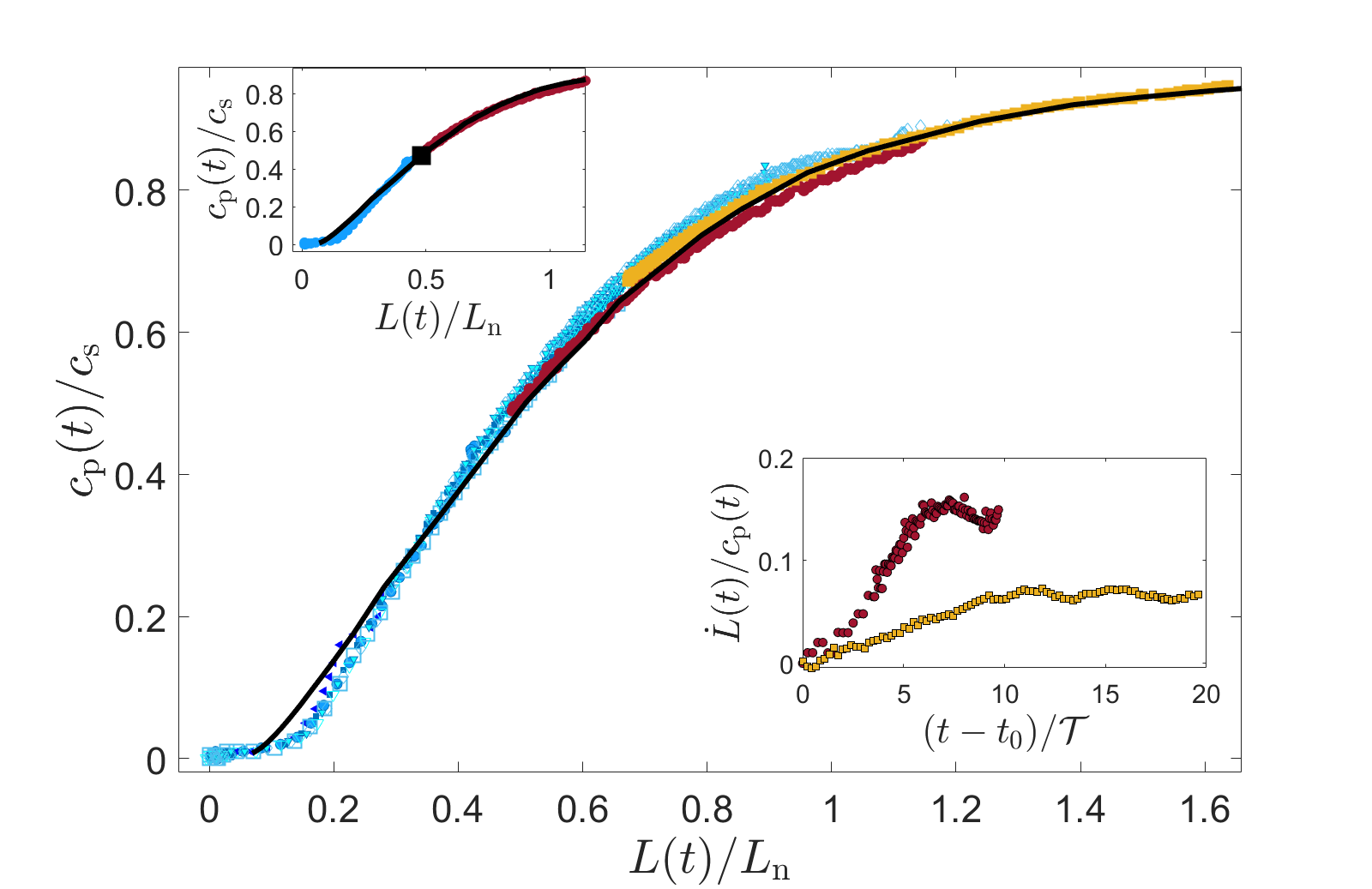}
\caption{(main) $L(t)\!-\!c_{\rm p}(t)$ trajectories corresponding to unsteady slip pulses generated from perturbed steady-state pulse solutions (various $\tau_{\rm d}$ and $\epsilon$ perturbation values). It is observed that under all conditions, unsteady pulses (whether growing, marked in hot colors, or decaying, marked in cold colors) closely follow the steady-state $c^{\mbox{\tiny{(0)}}}_{\rm p}[L^{\mbox{\tiny{(0)}}}]$ line (solid black line), if they are initially near it, as predicted in~\eqref{eq:solution}. Shown are 9 datasets, with $\tau_{\rm d}/\tau_{\rm min}$ in the range $1.025\!-\!1.07$ and $\epsilon$ in the range $-0.41\!-\!0$~\cite{SM}. The results for $\tau_{\rm d}\!=\!1.05\tau_{\rm min}$ with $\epsilon\!=\!0$ (decaying pulse, blue squares) and $\epsilon\!=\!-0.3$ (growing pulse, brown squares) are shown in the upper inset. The steady-state $(L^{\mbox{\tiny{(0)}}}\!(\tau_{\rm d}),c^{\mbox{\tiny{(0)}}}_{\rm p}\!(\tau_{\rm d}))$ point (black square) indeed separates growing from decaying pulses, as predicted. (lower inset) $\dot{L}(t)/c_{\rm p}(t)$ vs.~$t\!-\!t_0$, verifying the prediction in ~\eqref{eq:slow} for growing pulses (see the scale of the y-axis).}
\label{fig:fig3}
\end{figure}

Since steady-state pulse solutions reside on the $c^{\mbox{\tiny{(0)}}}_{\rm p}[L^{\mbox{\tiny{(0)}}}]$ line (by definition), using perturbed steady-state pulse solutions as initial conditions is most suitable in the present context. We employ the same $\phi(x,t)$ field $\epsilon$-perturbation scheme used above for the dynamic stability analysis (cf.~Fig.~\ref{fig:fig2} and the discussion related to it). In the upper inset of Fig.~\ref{fig:fig3}, we present the results for two initial conditions for $\tau_{\rm d}\=1.05\tau_{\rm min}$ (a different value compared to Fig.~\ref{fig:fig2}), one (brown points) that features an up perturbation relative to the point $(L^{\mbox{\tiny{(0)}}}\!(\tau_{\rm d}),c^{\mbox{\tiny{(0)}}}_{\rm p}\!(\tau_{\rm d}))$ (large black square) and another (blue points) that features a down perturbation relative to it. It is observed that the former becomes a growing pulse (increasing $L(t)$ and $c_{\rm p}(t)$) and the latter becomes a decaying pulse (decreasing $L(t)$ and $c_{\rm p}(t)$), as predicted. Moreover, and crucially, in both cases the trajectory $L(t)\!-\!c_{\rm p}(t)$ closely follows the steady-state $c^{\mbox{\tiny{(0)}}}_{\rm p}[L^{\mbox{\tiny{(0)}}}]$ line (solid black line) {\em at any time}, as predicted theoretically in ~\eqref{eq:solution}.

Next, in the main panel of Fig.~\ref{fig:fig3}, we superimpose $L(t)\!-\!c_{\rm p}(t)$ trajectories for pulses obtained from steady-state solutions at various $\tau_{\rm d}$ values and various $\epsilon$ perturbations (see figure caption, and note that the selected $\tau_{\rm d}$ values are such that expanding crack-like rupture is not excited~\cite{Brener2018}). Growing pulses appear in hot colors and decaying ones in cold colors. It is observed that independently of the value of $\tau_{\rm d}$ and independently of whether the resulting pulse is growing or decaying, all unsteady pulse trajectories reside on the steady-state $c^{\mbox{\tiny{(0)}}}_{\rm p}[L^{\mbox{\tiny{(0)}}}]$ line (solid black line) {\em at any time}, as predicted theoretically in ~\eqref{eq:solution}. Finally, in the lower inset we test the slow unstable growth prediction for growing pulses ($\dot{L}(t)\!>\!0$, hot colors). It is observed that $\dot{L}(t)/c_{\rm p}(t)\!\ll\!1$ during the entire pulse propagation, as expected according to ~\eqref{eq:slow}. These results demonstrate that growing pulses along the $c^{\mbox{\tiny{(0)}}}_{\rm p}[L^{\mbox{\tiny{(0)}}}]$ line are indeed sustained pulses.

In the second test of the theory, an even more stringent test, we aim at generating slip pulses that are initially arbitrarily far away from the steady-state $c^{\mbox{\tiny{(0)}}}_{\rm p}[L^{\mbox{\tiny{(0)}}}]$ line. This will allow us to test the prediction that pulse trajectories in the $L(t)\!-\!c_{\rm p}(t)$ plane are attracted to the $c^{\mbox{\tiny{(0)}}}_{\rm p}[L^{\mbox{\tiny{(0)}}}]$ line, and once they hit it (say at time $t_0$) they remain close to the line at later times. Moreover, the resulting pulses are predicted to be growing or decaying depending on the position of the hitting point $(L(t_0), c_{\rm p}(t_0))$ relative to $(L^{\mbox{\tiny{(0)}}}(\tau_{\rm d}),c^{\mbox{\tiny{(0)}}}_{\rm p}\!(\tau_{\rm d}))$. The latter two predictions have already been verified in Fig.~\ref{fig:fig3} for slip pulses that are initially close to the $c^{\mbox{\tiny{(0)}}}_{\rm p}[L^{\mbox{\tiny{(0)}}}]$ line.

Slip pulses that are initially arbitrarily far from the $c^{\mbox{\tiny{(0)}}}_{\rm p}[L^{\mbox{\tiny{(0)}}}]$ line are generated by introducing perturbations to a nearly quiescent frictional state (a state that slides at $v_{\rm stick}$, i.e.~at an extremely slow rate). The perturbations are initially stationary (i.e.~non-propagating), and may correspond to remote triggering processes and/or some interfacial heterogeneity. By varying the size of the perturbation, as explained in detail in~\cite{SM}, slip pulses of various properties can be generated.
\begin{figure}[ht!]
\centering
\includegraphics[width=1\columnwidth]{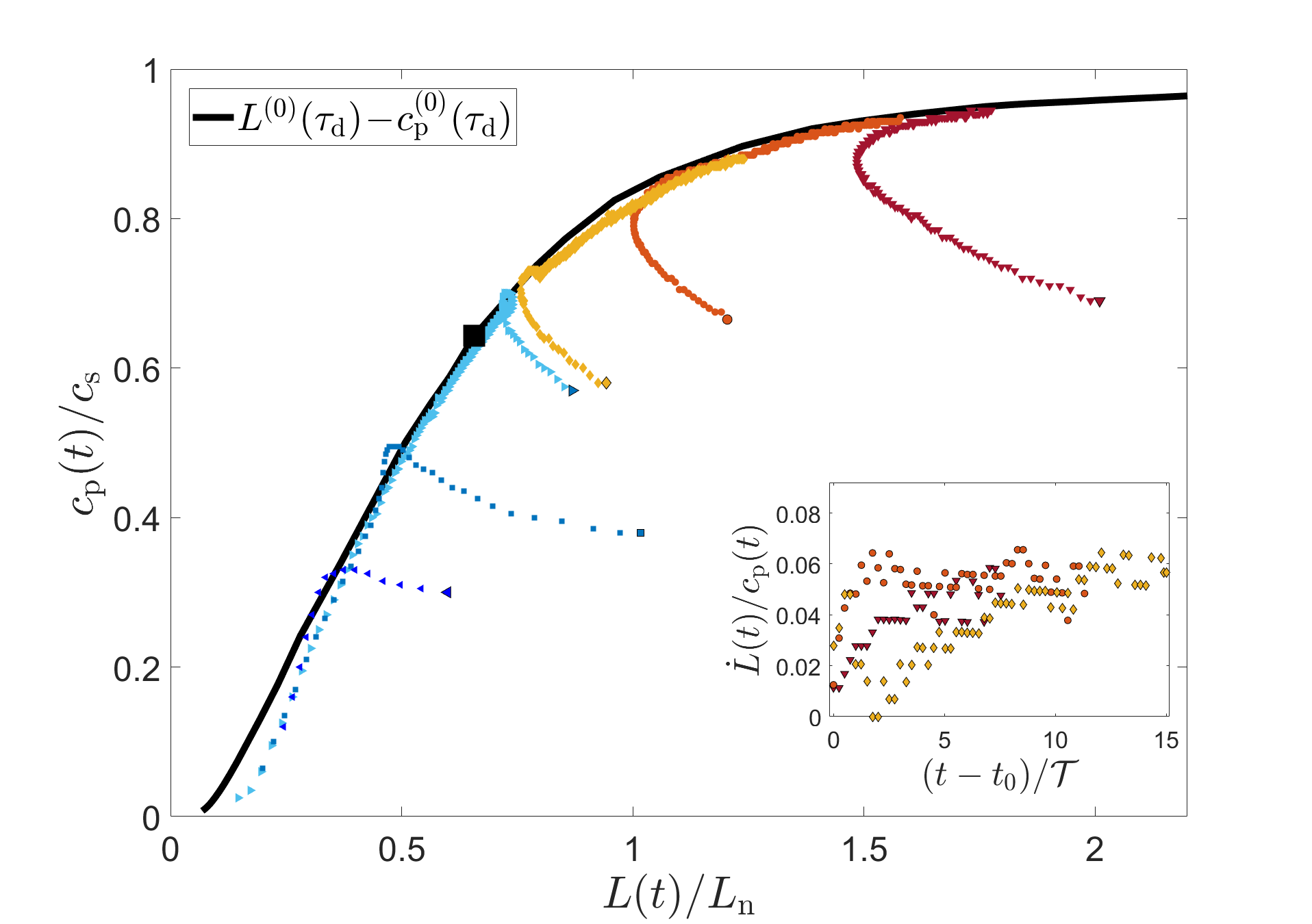}
\caption{The same as Fig.~\ref{fig:fig3}, but for 6 slip pulses that are initially far from the $c^{\mbox{\tiny{(0)}}}_{\rm p}[L^{\mbox{\tiny{(0)}}}]$ line (solid black line), under a fixed $\tau_{\rm d}\!=\!1.04\tau_{\rm min}$. Slip pulses are excited as detailed in the text and in~\cite{SM}, and their first time point is marked by a larger symbol in each case. It is observed that all pulse trajectories are attracted to the $c^{\mbox{\tiny{(0)}}}_{\rm p}[L^{\mbox{\tiny{(0)}}}]$ line, and once hitting the line (at a time denoted as $t_0$), they remain close to it at any $t\!>\!t_0$ time. Moreover, the steady-state $(L^{\mbox{\tiny{(0)}}}\!(\tau_{\rm d}),c^{\mbox{\tiny{(0)}}}_{\rm p}\!(\tau_{\rm d}))$ point (black square) separates growing (hot colors) from decaying (cold colors) pulses, as predicted. The prediction in ~\eqref{eq:slow} for growing pulses is verified in the inset (see the scale of the y-axis).}
\label{fig:fig4}
\end{figure}

In Fig.~\ref{fig:fig4}, we present results for 6 different pulse trajectories in the $L(t)\!-\!c_{\rm p}(t)$ plane for $\tau_{\rm d}\=1.04\tau_{\rm min}$. The first time at which well-defined slip pulse size $L$ and propagation velocity $c_{\rm p}$ exist in each case is marked by a larger symbol. It is observed that while all pulse trajectories are initially far from the $c^{\mbox{\tiny{(0)}}}_{\rm p}[L^{\mbox{\tiny{(0)}}}]$ line, they are dynamically attracted to it. Interestingly, the time it takes each pulse from formation to hitting the line is of order ${\cal T}(\tau_d)$~\cite{SM}, indicating that the latter is a basic timescale in the pulse problem in a way that goes beyond the linearized dynamics of ~\eqref{eq:L_dynamics_linear}. Once each trajectory hits the line (at time $t_0$), it remains close to it at any time. Moreover, the position of the hitting point relative to $(L^{\mbox{\tiny{(0)}}}\!(\tau_{\rm d}),c^{\mbox{\tiny{(0)}}}_{\rm p}\!(\tau_{\rm d}))$ (large black square) approximately determines whether the dynamics along the line correspond to decaying pulses (cold colors) or to growing pulses (hot colors). For the latter, we find that $\dot{L}(t)/c_{\rm p}(t)\!\ll\!1$ for $t\!>\!t_0$ (see inset of Fig.~\ref{fig:fig4}), as expected according to ~\eqref{eq:slow}, demonstrating again that the growing pulses are indeed sustained pulses.

{\em Summary and discussion}.--- A comprehensive theoretical picture of the dynamics of unsteady frictional slip pulses has been developed. We showed that steady-state slip pulses are intrinsically unstable objects, yet the steady-state line $L(\tau_{\rm d})\!-\!c_{\rm p}(\tau_{\rm d})$ (parameterized by the driving stress $\tau_{\rm d}$) is of fundamental importance for unsteady pulse dynamics. In particular, while each point along the $c^{\mbox{\tiny{(0)}}}_{\rm p}[L^{\mbox{\tiny{(0)}}}]$ line is unstable, the line itself is a dynamic attractor for pulses under a given $\tau_{\rm d}$.

Once a general pulse $L(t)\!-\!c_{\rm p}(t)$ trajectory hits the $c^{\mbox{\tiny{(0)}}}_{\rm p}[L^{\mbox{\tiny{(0)}}}]$ line, it remains close to it at any time. The  position of the hitting point relative to $(L^{\mbox{\tiny{(0)}}}\!(\tau_{\rm d}),c^{\mbox{\tiny{(0)}}}_{\rm p}\!(\tau_{\rm d}))$ determines whether the dynamics along the line correspond to decaying or growing pulses. Finally, pulse dynamics along the $c^{\mbox{\tiny{(0)}}}_{\rm p}[L^{\mbox{\tiny{(0)}}}]$ line are controlled by a single slow unstable mode, which for growing pulses is manifested by $\dot{L}(t)/c_{\rm p}(t)\!\ll\!1$. The latter corresponds to sustained pulses, which are of prime importance for the failure dynamics of frictional systems~\cite{Brener2018}. The comprehensive theoretical picture has been quantitatively supported by large-scale, dynamic boundary-integral method simulations. We stress that we did not discuss here the conditions for exciting slip pulses (rather than expanding crack-like rupture), an issue that has been previously discussed in~\cite{Zheng1998,Gabriel2012,Brener2018}, but rather focused on their unsteady dynamics once they are excited.

The emerging theory suggests that the knowledge of the family of steady-state pulse solutions, in particular the steady-state line $L(\tau_{\rm d})\!-\!c_{\rm p}(\tau_{\rm d})$, allows to understand and predict many of the properties of unsteady slip pulses, when they are excited. The deep relation between unsteady slip pulses and their steady-state counterparts is valid for other discrete, coarse-grained variables (such as $L(t)$ and $v_{\rm m}(t)$) and in fact also for the entire $v(x,t)$ and $\phi(x,t)$ fields themselves, as shown in~\cite{SM}. The theory also predicts the linearized dynamics near each unstable point, cf.~~\eqref{eq:L_dynamics_linear}, with a characteristic timescale ${\cal T}(\tau_{\rm d})\!\sim\!L^{\mbox{\tiny{(0)}}}(\tau_{\rm d})/c^{\mbox{\tiny{(0)}}}_{\rm p}(\tau_{\rm d})$. In fact, the timescale ${\cal T}(\tau_{\rm d})$ also characterizes the approach to the $c^{\mbox{\tiny{(0)}}}_{\rm p}[L^{\mbox{\tiny{(0)}}}]$ line. Hence, ${\cal T}(\tau_{\rm d})$ ---  which is also obtained from steady-state pulse solutions --- appears to play the role of a basic timescale in the problem.

In addition, while the theory does not separately predict the general nonlinear dynamics of $\dot{L}(t)$ and $\dot{c}_{\rm p}(t)$ along the line, their ratio is given as $\dot{L}(t)/\dot{c}_{\rm p}(t)\=(dc^{\mbox{\tiny{(0)}}}_{\rm p}[L(t)]/dL(t))^{-1}$. Finally, ~\eqref{eq:solution} may be viewed as an equation of motion for unsteady pulses, where their velocity $c_{\rm p}(t)$ instantaneously follows $L(t)$. This interpretation bears similarities to the equation of motion of ordinary cracks in bulk materials, where the crack's tip is viewed as a massless defect whose velocity instantaneously follows the crack length~\cite{Freund1998}.

The numerical support to the theory, as presented above, employed a generic rate-and-state dependent friction law and anti-plane symmetry conditions. We have no obvious reasons to expect any major/qualitative differences in the theoretical picture once different rate-and-state dependent friction laws (or frictional constitutive parameters) and in-plane symmetry conditions are considered. Yet, future work should test the emerging picture in a wider range of physical situations. Interestingly, the results presented in~\cite{brantut2019stability} for a thermal pressurization constitutive relation indicate that unsteady slip pulses (when they are excited, and not expanding crack-like rupture) are related to the corresponding family of steady-state pulse solutions, similarly to our findings. The thermal pressurization results also indicate the relevance of the emerging picture to strongly weakening frictional interfaces, as well as to interfaces in which self-healing pulses do not depend on aging in stationary contact.

In general, knowledge of the interfacial constitutive relation (friction law) may allow one to come up with concrete predictions for unsteady pulse dynamics, which can be experimentally tested (e.g., in laboratory experiments). Future work should also address extensions of the present theory to inhomogeneous driving stresses.

Finally, in a broader context, the theory of the dynamics of unsteady slip pulses in frictional systems contributes to our general understanding of the dynamics of nonlinear spatially-extended dissipative systems. This particularly pertains to the reduced-dimensionality description in terms of a two coarse-grained dynamical variables, such as $L(t)$ and $c_{\rm p}(t)$, whose dynamics are controlled by a single slow unstable mode. Such a theoretical picture bears some similarities to the dynamics of equilibrium first-order phase transitions, such as solidification~\cite{lifschitz1983physical}, involving the formation of critical nuclei and front propagation in which a stable phase invades an unstable one (e.g.~crystal growth in the solidification problem). In such cases, the dynamics are also dominated by a single slow unstable mode, which corresponds to the size of the growing nucleus.

\vspace{0.2cm}
{\em Author contributions}\, E.B.~and E.A.B. designed the research, conceptualized the research, developed the theory and guided the numerical simulations, A.P.~performed all numerical simulations and their analysis, F.B.~and T.R.~helped with the numerical simulations, E.B.~and A.P.~wrote the manuscript.

\vspace{0.2cm}
{\em Acknowledgements} This work has been supported by the Israel Science Foundation (grant no.~1085/20). E.B.~acknowledges support from the Ben May Center for Chemical Theory and Computation and the Harold Perlman Family.

\clearpage

\onecolumngrid
\begin{center}
	\textbf{\large Supplemental Materials for: ``The dynamics of unsteady frictional slip pulses''}
\end{center}

\setcounter{equation}{0}
\setcounter{figure}{0}
\setcounter{section}{0}
\setcounter{table}{0}
\setcounter{page}{1}
\makeatletter
\renewcommand{\theequation}{S\arabic{equation}}
\renewcommand{\thefigure}{S\arabic{figure}}
\renewcommand{\thesection}{S-\arabic{section}}
\renewcommand{\thetable}{S-\arabic{table}}
\renewcommand*{\thepage}{S\arabic{page}}

\twocolumngrid

The goal of this document is to provide some technical details that allow to reproduce the results presented in the manuscript and to offer some additional supporting data.

\section{The interfacial constitutive relation (friction law)}

The coupled bulk-interface frictional problem we solved is formulated using the boundary integral approach, where the shear stress at the interface (located at $y\=0$) is given on the right-hand-side (RHS) of~\cite{das1980numerical}
\begin{equation}
\label{eq:BIM}
\tau[v(x,t),\phi(x,t)] = \tau_{\rm d} -\frac{\mu}{2c_s}v(x,t) + s(x,t) \ ,
\end{equation}
as presented in Eq.~(1) in the manuscript. Here, $\tau_{\rm d}$ is the driving stress applied far from the interface and $(\mu/2c_{\rm s})v(x,t)$ ($\mu$ and $c_{\rm s}$ are the shear modulus and wave-speed of the bulks, respectively) is the so-called radiation damping term, which depends on the local slip velocity $v(x,t)$. $s(x,t)$ is a spatiotemporal convolutional integral that accounts for the long-range interaction of different parts of the interface, mediated by bulk deformation. In general, $s(x,t)$ does not admit an explicit real-space representation~\cite{Breitenfeld1998}, but is rather expressed in terms of the Fourier components of $v(x,t)$ (in fact, it involves the slip displacement $u(x,t)\=\int\! v(x,t)\,dt$, obtained from $v(x,t)$ through time integration). Its specific form is different for anti-plane shear symmetry (i.e.~slip displacement in the $z$ direction), which is used in our computer simulations, and for in-plane shear symmetry (i.e.~slip displacement in the $x$ direction).

The interfacial shear stress (RHS of Eq.~\eqref{eq:BIM}) is balanced at each point in time and interfacial position by the frictional strength $\tau[v(x,t),\phi(x,t)]$, which depends in addition to $v(x,t)$ on the internal state field $\phi(x,t)$ (see manuscript for discussion). The frictional strength we employed, which is based on extensive laboratory data~\cite{Marone1998a,Baumberger2006Solid,Bar-Sinai2014}, takes the form~\cite{Brener2018}
\begin{eqnarray}
\label{eq:f}
\tau(v,\phi)/\sigma &=& \left[1+b\log\left(1+\phi/\phi^{\star}\right)\right] \times \\ \nonumber
&&\left[f_0/\sqrt{1+\left(v^{\star}\!/v\right)^2}+\alpha\,\log\left(1+{\abs v}/v^{\star}\right)\right]\ .
\end{eqnarray}
The frictional parameters appearing in Eq.~\eqref{eq:f} are given in Table~\ref{t:1}, along with the linear elastic bulk parameters of Eq.~\eqref{eq:BIM} and the value of the normal stress $\sigma$~\cite{Brener2018}. As explained in the manuscript, the internal state field $\phi(x,t)$ satisfies a well-established local (in space) differential equation (known as the `aging-law'~\cite{Marone1998a,Baumberger2006Solid})
\begin{equation}
\label{eq:phi}
\partial_t\phi(x,t)\=1-\!\frac{v(x,t)\,\phi(x,t)}{D}\, .
\end{equation}
The latter include the characteristic slip distance $D$, whose value is also given in Table~\ref{t:1}.
\begin{table}
\begin{tabular}{|c|c |c|}
\hline
$\sigma$ & $1$  & MPa\\
	\hline
$\mu$            & $9$&	GPa\\
\hline
$c_{\rm s} $      &$2739$  &m/s\\
\hline
 $D$              &$5\times 10^{-7}$ & m\\
 \hline
  $v^*$           &$10^{-7}$  &m/s\\
  \hline
$\phi^*$         &$3.3\times10^{-4}$ & s\\
\hline
$\alpha$         & $0.005$	&--\\
\hline
$b$              &$0.075$  &--\\
\hline
  $f_0$          &$0.28$ & --\\
	\hline
\end{tabular} \label{t:1}
\caption{The normal stress $\sigma$, linear elastic bulk parameters and frictional parameters (cf.~Eqs.~\eqref{eq:BIM}-\eqref{eq:f}) used in the computer simulations.}
\end{table}

Under persistent sliding at a slip velocity $v$, Eq.~\eqref{eq:phi} attains a steady state of the form $\phi\=D/v$. When the latter is substituted in $\tau(v,\phi)$ of Eq.~\eqref{eq:f}, the steady-state friction curve of Fig.~1 in the manuscript (solid brown line therein) is obtained. It is generally N-shaped, featuring a local minimum at $(v_{\rm min}, \tau_{\rm min})$ (cf.~Fig.~1 in the manuscript). While the local minimum emerges from basic physical considerations and is supported by experimental data~\cite{Baumberger2006Solid,Bar-Sinai2014}, its existence does not play a fundamental role in the slip pulses problem we solved. Yet, the characteristic slip velocity  $v_{\rm min}\=0.005971$ m/s and characteristic frictional strength $\tau_{\rm min}\=0.3406679$ MPa can be used to nondimensionalize various quantities, as done in the manuscript. In particular, the driving stress we employed was varied in the range $\tau_{\rm d}/\tau_{\rm min}\!\in\! [1.015, 1.07]$. Moreover, using the previously-calculated effective fracture energy $G_{\rm c}\=0.65$~J/m$^2$ of slip pulses for the frictional parameters in Table~\ref{t:1}~\cite{pomyalov2023self}, one can construct a characteristic pulse length as $L_{\rm n}\=(c_{\rm s}/v_{\rm min})^2 (G_{\rm c}/\mu)\=15.2$~m. The latter is used to normalize quantities of length dimension in the manuscript.

\section{Dynamic boundary-integral method simulations using the open-source library cRacklet}

The mathematical problem defined by Eqs.~\eqref{eq:BIM}-\eqref{eq:phi} is solved using the open-source library {\em cRacklet}~\cite{roch2022cracklet}. The latter is based on a spectral formulation of linear elastodynamics for two semi-infinite solids  based on the relevant Green's function (cf.~RHS of Eq.~\eqref{eq:BIM})~\cite{Breitenfeld1998}. Its implementation is specially tailored for  modeling dynamic rupture propagation along a planar interface bonding the two half-spaces. In particular, it allows to implement an interfacial constitutive relation $\tau[v(x,t),\phi(x,t)]$ (cf.~LHS of Eq.~\eqref{eq:BIM} and Eq.~\eqref{eq:f}) and accompanying differential equations such as Eq.~\eqref{eq:phi}.

The spectral formulation, which involves a Fourier representation of the relevant fields, implies that all fields are periodic in the spatial integration domain of size $W$. That is, in our case we have $v(x,t)\=v(x+W,t)$ and $\phi(x,t)\=\phi(x+W,t)$ at any time $t$. Moreover, initial conditions for the two fields should be provided, i.e.~$v(x,t\=0)$ and $\phi(x,t\=0)$. Finally, past frictional dynamics (i.e.~for $t\!<\!0$) can be incorporated through the slip displacement history $u(x,t\!\le\!0)$. The latter implies a non-vanishing contribution to the interfacial stress at $t\=0$, $s(x,t\=0)$. Next, we discuss the two types of initial conditions employed in our numerical simulations using {\em cRacklet}.


\section{Two types of initial conditions}

In the manuscript, two types of initial conditions are employed, depending on the physical question of interest. Here, we briefly describe both.

\subsection{Initial conditions based on steady-state slip pulses}
\label{subsec:IC_ss}

A family of steady-state slip pulse solutions, parameterized by the driving stress $\tau_{\rm d}$, has been recently obtained numerically in~\cite{pomyalov2023self}. Such solutions take the form $v_{_{\rm ss}}\!(\xi)$ and $\phi_{_{\rm ss}}\!(\xi)$, with $\xi\!=\!x-c^{\mbox{\tiny{(0)}}}_{\rm p}t$ being a co-moving space-time coordinate and $c^{\mbox{\tiny{(0)}}}_{\rm p}$ is the steady pulse propagation velocity. For reasons that are explained in detail in the manuscript, one is interested in using these steady-state pulse solutions (or a slightly perturbed version of them, see below) as initial conditions to the set of integro-differential equations defined in Eqs.~\eqref{eq:BIM}-\eqref{eq:phi}.

To achieve this, we proceed in two steps. First, one needs to conform with the periodic boundary conditions employed in solving Eqs.~\eqref{eq:BIM}-\eqref{eq:phi} using {\em cRacklet}. In principle, steady-state slip pulse solutions correspond to a closed (homoclinic) orbit $(v_{_{\rm ss}}\!(\xi),\phi_{_{\rm ss}}\!(\xi))$ in the $v-\phi$ plane (the corresponding closed orbit in the $v-\tau$ plane is presented in Fig.~1 in the manuscript, cf.~solid green line therein). Consequently, these solutions essentially satisfy periodic boundary conditions. Yet, since slip pulses feature long, non-exponential tails~\cite{pomyalov2023self}, the actual numerical solutions may not strictly correspond to closed orbits. Whenever a difference in the $v$ and $\phi$ values at both ends of a steady-state pulse solution exists, we imposed periodicity by exponentially extrapolating the fields between the different values at the tail (we checked that a linear extrapolation for the velocity field does not affect the results for large enough $W$) and by padding the values ahead of the pulse to be accommodated into the {\em cRacklet} numerical grid of overall size $W$. Interpolation is then invoked to assign values of the fields at each grid point.

\begin{figure}[htp]
\includegraphics[width=1.05\columnwidth]{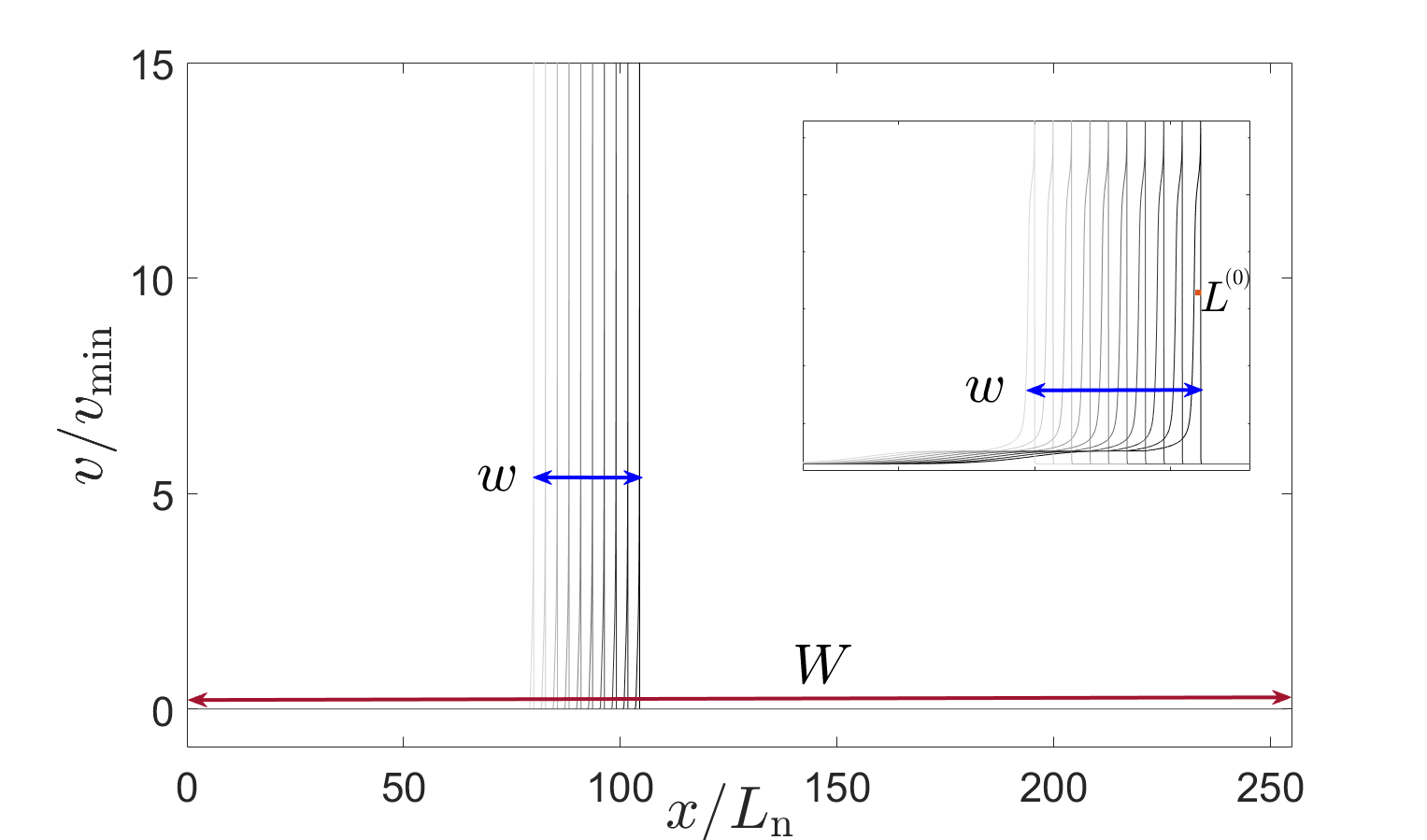}
	\caption{An illustration of the procedure by which slip history based on steady-state slip pulses is generated. A system with spatial periodicity $W$ in the $x$ direction is considered. A steady-state pulse solution $v_{_{\rm ss}}\!(x-c^{\mbox{\tiny{(0)}}}_{\rm p}t)$ is placed at $t\!=\!0$ not close to the periodic boundaries of the system (see the rightmost field in the main panel and note that the $y$-axis is truncated for better visibility). Moving time backwards (i.e.~into the past, $t\!<\!0$), $v_{_{\rm ss}}\!(x-c^{\mbox{\tiny{(0)}}}_{\rm p}t)$ is ``dragged'' in space according to its propagation velocity $c^{\mbox{\tiny{(0)}}}_{\rm p}$, representing steady-state pulse propagation in the time interval $-w/c^{\mbox{\tiny{(0)}}}_{\rm p}\!<\!t\!<\!0$. The spatial extent of steady-state propagation $w$ is selected to be much smaller than $W$, i.e.~$w\!\ll\!W$. The latter ensures that the boundary on the left ($x\!=\!0$) is not reached. Had it reached the left boundary, the pulse would have emerged from the right boundary (due to the spatial periodicity), implying a spurious future effect, which must be avoided. In addition, to mimic steady-state pulse propagation in the past, $w$ has to be much larger than $L^{\mbox{\tiny{(0)}}}$, i.e.~$w\!\gg\!L^{\mbox{\tiny{(0)}}}$. The latter is illustrated in the inset, where the slip velocity snapshots in the main panel are re-plotted on a semi-logarithmic scale, which allows to mark $L^{\mbox{\tiny{(0)}}}$ by the tiny orange marker (corresponding to    $L^{\mbox{\tiny{(0)}}}/L_{\rm n}\!=\!0.96$). Overall, the figure illustrates the requirement  $L^{\mbox{\tiny{(0)}}}\!\ll\!w\!\ll\!W$, as discussed in Sect.~\ref{subsec:IC_ss}.}
 \label{f:hist}
\end{figure}

Second, with $W$-periodic steady-state pulse solutions at hand, one needs to generate slip history that corresponds to steady-state pulses that propagated steadily at a velocity $c^{\mbox{\tiny{(0)}}}_{\rm p}$ indefinitely in the past, $-\infty\!<\!t\!<\!0$. This condition, however, is clearly in conflict with the spatial periodic boundary conditions used. That is, extending the history (past propagation) beyond the domain boundary implies that the past spuriously affects future dynamics. Therefore, we require that the past propagation distance of the steady-state pulse, denoted by $w$, is much smaller than the spatial periodicity domain $W$, $w\!\ll\!W$, see Fig.~\ref{f:hist}. Moreover, the position of the pulse at $t\=0$ should be also sufficiently far from the domain boundary, as is demonstrated in Fig.~\ref{f:hist} as well.

On the other hand, to approximately mimic steady-state propagation, $w$ has to be much larger than the characteristic slip pulse size $L^{\mbox{\tiny{(0)}}}$. The latter is defined, following~\cite{pomyalov2023self}, as the width at the half-height of a logarithmic representation of the slip velocity field, corresponding to the solutions of $\log(1+v(\xi)/v^*)/\log(1+v_{\rm m}/v^*)\=1/2$. Here, $v_{\rm m}$ is the maximal slip velocity of the pulse, cf.~the inset of Fig.~1 in the manuscript. This operative pulse size definition is uniformly applied in the entire manuscript, also to time-dependent unsteady pulses, resulting in $L(t)$ (the corresponding pulse propagation velocity $c_{\rm p}(t)$ is determined by the time rate of change of the spatial position of $v_{\rm m}(t)$). Different operational definitions of the pulse width did not affect any of the reported results. Consequently, past steady-state pulse propagation can be mimicked by choosing $w\!\gg\!L^{\mbox{\tiny{(0)}}}$, see Fig.~\ref{f:hist}.

Taken together, using initial conditions based on steady-state slip pulses entails the resolution of the following condition $L^{\mbox{\tiny{(0)}}}\!\ll\!w\!\ll\!W$, which is computationally demanding (see additional discussion of this point below). Once a given slip history is generated, it is used to calculate the spatiotemporal convolutional integral and the resulting field $s(x,t\=0)$ is used in Eq.~\eqref{eq:BIM} as an initial condition. The latter can be supplemented with a perturbation in the initial internal state field, $\phi(x,t\=0)$, as will be further discussed below.

\subsection{Stationary perturbations of a nearly quiescent frictional state}
\label{subsec:skewed_Gaussian}

Another form of initial condition employed in the manuscript is obtained by perturbing at $t\=0$ a nearly quiescent frictional state. The latter corresponds to a spatially homogeneous frictional state at a slip velocity $v_{\rm stick}$, which is the smallest $v$ solution to the equation $\tau[v,\phi\=D/v]\=\tau_{\rm d}$ (cf.~Fig.~1 in the manuscript). Since $v_{\rm stick}$ is an extremely small slip velocity, the corresponding frictional state is essentially quiescent (i.e.~non-sliding/sticking).

At $t\=0$, we introduce a slip velocity perturbation such that $v(x,t\!=\!0)$ takes the form
\begin{eqnarray}
\label{eq:skew}
&\hspace{-5cm} v(x,t\!=\!0)- v_{\rm stick}= \, \\ \nonumber
& \,\, \Delta{v}\left(1 + {\rm erf}\!\left[s\,\frac{\displaystyle (x-x_0)}{ \displaystyle \sqrt{2}\,\Delta{L}}\right]\right) \exp\!\left[-\frac{\displaystyle (x-x_0)^2}{\displaystyle 2\,(\Delta{L})^2}\right] \, ,\\ \nonumber
\end{eqnarray}
where $\Delta{v}$ is the amplitude of the perturbation, $x_0$ is the position of the maximum of the perturbation, $\Delta{L}$ is the width of the perturbation and $s$ is the dimensionless skewness/shape parameter. We used the skewed/asymmetric Gaussian perturbations of Eq.~\eqref{eq:skew} so as to favor the generation of a single slip pulse (while symmetric perturbations may give rise to a pair of pulses~\cite{Brener2018}). We set $\Delta{v}\=5\times 10^{-4}$ m/s and $s\=5$ in all of the calculations reported in the manuscript in relation to Fig.~4 therein. We found that for a range of perturbation widths $\Delta{L}$, the above-described initial condition leads to the excitation of slip pulses (see additional information in Fig.~\ref{f:vmax} below).

\section{Dynamic evolution of unstable steady-state pulses}

As explained at length in the manuscript, the first application of the initial conditions described in Sect.~\ref{subsec:IC_ss} is in studying the dynamic stability of steady-state pulses using {\em cRacklet}. To that aim, as is also explained in the manuscript, we supplemented the slip history generated based on steady-state pulse solutions (as detailed in Sect.~\ref{subsec:IC_ss}) with perturbations of the $\phi$ field at $t\=0$. Specifically, after the slip history is generated, we introduced $\phi(x,t\=0^-)$ (also corresponding to the steady-state pulse solution) with an amplitude perturbation of the form $\phi(x,t\=0)\=(1+\epsilon)\phi(x,t\=0^-)$, quantified by $\epsilon$.

Performing dynamic stability analysis of steady-state pulses amounts to using small values of $|\epsilon|$ and following the dynamics in {\em cRacklet} over relatively short periods of time $t\!>\!0$. Moreover, as explained in the manuscript, we theoretically expect steady-state slip pulses to be categorically unstable, where $\epsilon\!<\!0$ leads to growing pulses and $\epsilon\!>\!0$ leads to decaying pulses. That is, we theoretically expect a critical $\epsilon_{\rm c}\!\approx\!0$ to separate growing and decaying pulses.

A very accurate test of this theoretical expectation is somewhat challenging in view of the requirement $L^{\mbox{\tiny{(0)}}}\!\ll\!w\!\ll\!W$, due to the finite system size, and the computational limitations associated with it. The quality of resolution of this condition determines the degree by which we represent steady pulse propagation in the past ($t\!<\!0$), and hence is expected to affect the quantitative value of $\epsilon_{\rm c}$. For any reasonable resolution of this condition, we found that indeed steady-state pulses are categorically unstable and that there exists a critical $\epsilon_{\rm c}$ that separates growing and decaying pulses, as expected theoretically.

The actual value of $\epsilon_{\rm c}$ was found to be finite and negative. To demonstrate that it is indeed a finite size effect, we extracted $\epsilon_{\rm c}$ for calculations in which we systematically improved the resolution of the condition $L^{\mbox{\tiny{(0)}}}\!\ll\!w\!\ll\!W$. We found that, as expected, $|\epsilon_{\rm c}|$ decreases when $L^{\mbox{\tiny{(0)}}}\!\ll\!w\!\ll\!W$ is better resolved. In the manuscript, we show (cf.~Fig.~2 therein) that for $L^{\mbox{\tiny{(0)}}}/L_{\rm n}\=0.96$, $w/L_{\rm n}\!\approx\!25$ and $W/L_{\rm n}\!\approx\!255$, we have $\epsilon_{\rm c}\!\approx\!-0.1$. Further reducing $\epsilon_{\rm c}$ would necessitate very long calculations, which are not needed in view of the trend of convergence of $\epsilon_{\rm c}$ toward zero.

\begin{figure}[htp]
\includegraphics[width=1.05\columnwidth]{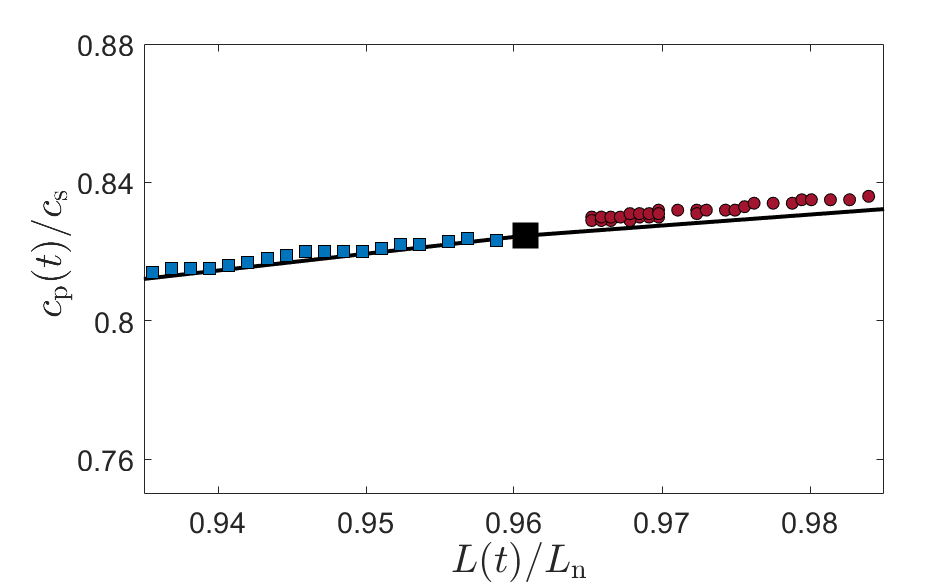}
	\caption{\label{f:tangent1.03}  The data shown in Fig.~2a-b in the manuscript for $L(t)$ and $c_{\rm p}(t)$ are re-plotted (for a growing pulse and a decaying pulse, the same colors and symbols) in the $L\!-\!c_{\rm p}$ plane, parameterized by $t$ for short times. Superimposed are the steady-state $c^{\mbox{\tiny{(0)}}}_{\rm p} [L^{\mbox{\tiny{(0)}}}]$ line (solid black line) and the steady-state $(L^{\mbox{\tiny{(0)}}}(\tau_{\rm d}), c^{\mbox{\tiny{(0)}}}_{\rm p}(\tau_{\rm d}))$ point (black square). It is observed, as predicted, that the two pulse trajectories evolve along the local tangent to the steady-state line at that point.}	
\end{figure}

Additional support to the expected unstable nature of slip pulses is given in the inset of Fig.~2a in the manuscript, where the linearized dynamics of Eq.~(3) therein are verified. Yet another strong supporting evidence for the theoretical predictions and their numerical validation is provided in Fig.~\ref{f:tangent1.03}, where the short time dynamics of $L(t)$ and $c_{\rm p}(t)$ of Fig.~2a-b in the manuscript, for both growing (brown data) and decaying (blue data), are plotted in the $L\!-\!c_{\rm p}$ plane, superimposed on the steady-state $c^{\mbox{\tiny{(0)}}}_{\rm p}[L^{\mbox{\tiny{(0)}}}]$ line. It is observed that the two pulse trajectories initiate close to the corresponding steady-state $(L^{\mbox{\tiny{(0)}}}(\tau_{\rm d}), c^{\mbox{\tiny{(0)}}}_{\rm p}(\tau_{\rm d}))$ point (black square) and evolve along the local tangent to the steady-state $c^{\mbox{\tiny{(0)}}}_{\rm p}[L^{\mbox{\tiny{(0)}}}]$ line at that point. As discussed in the manuscript, this result supports the expectation that the single unstable mode in the problem is directed along the local tangent.

A second application of the initial conditions described in Sect.~\ref{subsec:IC_ss} is in generating slip pulses that are initially relatively close to their corresponding steady-state $(L^{\mbox{\tiny{(0)}}}\!(\tau_{\rm d}), c^{\mbox{\tiny{(0)}}}_{\rm p}\!(\tau_{\rm d}))$ point (and to the steady-state $c^{\mbox{\tiny{(0)}}}_{\rm p}[L^{\mbox{\tiny{(0)}}}]$ line in general), though not insisting on being very close to it (as was the goal in the dynamic stability analysis discussed above). Therefore, we use a reasonable resolution of $L^{\mbox{\tiny{(0)}}}\!\ll\!w\!\ll\!W$ (though not a stringent one) and a wider range of $\epsilon$ values. In particular, in Fig.~3 in the manuscript we present results for $\tau_{\rm d}/\tau_{\rm min}\=1.05$ with $\epsilon\=0$ (blue squares) and $\epsilon\=-0.3$ (brown squares), shown also in the inset therein, for $\tau_{\rm d}/\tau_{\rm min}\=1.07,\,\epsilon\=0$ (filled blue left triangle), for $\tau_{\rm d}/\tau_{\rm min}\=1.04, \,\epsilon\=0$ (large open blue squares), for $\tau_{\rm d}/\tau_{\rm min}\=1.04,\,\epsilon\,=-0.3$ (small blue filled squares), for $\tau_{\rm d}/\tau_{\rm min}\=1.03, \,\epsilon\=0$ (light open  blue down triangles), for $\tau_{\rm d}/\tau_{\rm min}\=1.03,\,\epsilon\=-0.3$ (light blue  filled down triangles), for $\tau_{\rm d}/\tau_{\rm min}\=1.025,\,\epsilon\=0$ (light blue open diamonds), for $\tau_{\rm d}/\tau_{\rm min}\=1.04,\,\epsilon\=-0.41$ (filled yellow squares). All these simulations were carried out using $w/L_{\rm n}\!\approx\!15$.

\section{Additional supporting results}

Here, we present a few additional results that further support the theoretical picture of unsteady slip pulses developed in the manuscript.

\subsection{Dynamics in the $L\!-\!v_{\rm m}$ plane}

In Fig.~4 in the manuscript, we present the dynamics of unsteady slip pulses (generated through stationary perturbations of a nearly quiescent frictional state as explained in Sect.~\ref{subsec:skewed_Gaussian}) as trajectories in the $L\!-\!c_{\rm p}$ plane. This reduced-dimensionality description in terms of discrete, coarse-grained variables turned out to be most useful. In particular, the attraction of the various pulse trajectories to the steady-state $c^{\mbox{\tiny{(0)}}}_{\rm p}[L^{\mbox{\tiny{(0)}}}]$ line (which are initially far from it), the adherence of the various trajectories to the steady-state line after they hit it and the sorting of the different trajectories to growing/decaying pulses based on the position of the hitting point relative to the steady-state $(L^{\mbox{\tiny{(0)}}}(\tau_{\rm d}), c^{\mbox{\tiny{(0)}}}_{\rm p}(\tau_{\rm d}))$ point all provide great support to the developed theoretical picture of unsteady slip pulses.

\begin{figure}[htp]
\includegraphics[width=1\columnwidth]{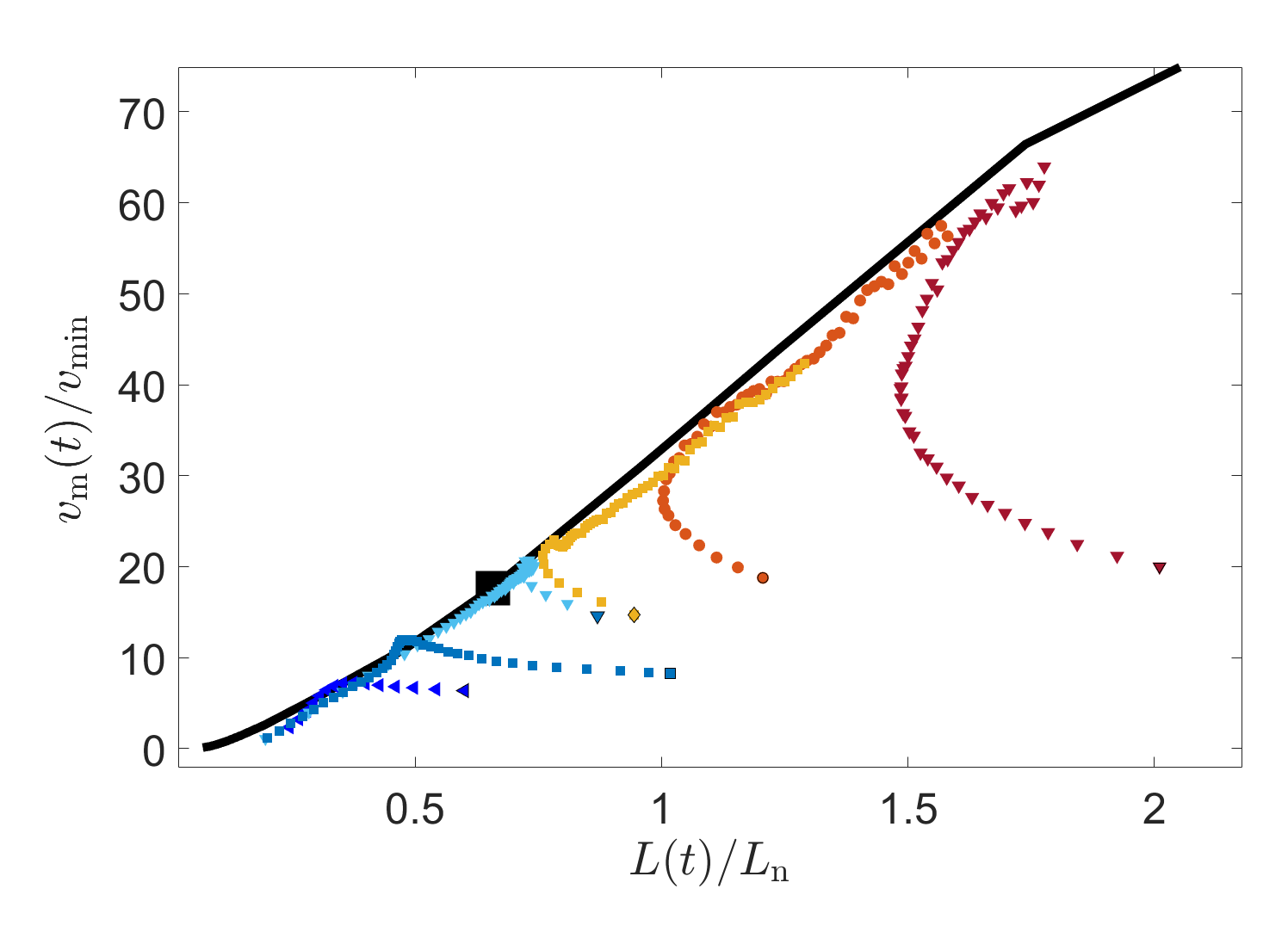}
\caption{\label{f:vmax} The same data as Fig.~4 in the manuscript (same colors and symbols), but for $v_{\rm m}(t)$ against $L(t)$ for each pulse. Pulses are generated using Eq.~\eqref{eq:skew} (see discussion around it) with perturbation widths with $\Delta{L}/L_{\rm n}\!=\!36.77,\,42.62,\,52.05,\,53.69,\,63.11,\,121.72$ (from bottom to top). The steady-state $v_{\rm m}^{\mbox{\tiny{(0)}}} [L^{\mbox{\tiny{(0)}}}]$ line (solid black line) and the steady-state $(L^{\mbox{\tiny{(0)}}}(\tau_{\rm d}), v_{\rm m}^{\mbox{\tiny{(0)}}}(\tau_{\rm d}))$ point (large black square) are added. The results are essentially the same as in Fig.~4 in the manuscript, see text for discussion. In addition, the time it takes each pulse --- from the moment it is formed to hitting the line --- is $0.62, 1.5,3.64, 1.57,2.32$ and $5{\cal T}$, respectively (again, from bottom to top). Indeed, as stated in the manuscript, this time is ${\cal O}({\cal T})$.}	
\end{figure}

Yet, the developed theoretical picture indicates that unsteady pulse dynamics are controlled by a single slow unstable mode and hence implies that while the choice of $L(t)$ and $c_{\rm p}(t)$, which is clearly (strongly) physically motivated, is not unique. That is, it implies that a pair of other discrete, coarse-grained variables could have revealed the same picture. To test this prediction, we use the very same simulations that gave rise to the results in Fig.~4 in the manuscript and present them in Fig.~\ref{f:vmax} in the $L\!-\!v_{\rm m}$ plane, along with the steady-state $v_{\rm m}^{\mbox{\tiny{(0)}}} [L^{\mbox{\tiny{(0)}}}]$ line (solid black line) and the steady-state $(L^{\mbox{\tiny{(0)}}}(\tau_{\rm d}), v_{\rm m}^{\mbox{\tiny{(0)}}}(\tau_{\rm d}))$ point. The results are essentially the same as in Fig.~4 in the manuscript, lending additional support to the theory. Note that the values of the perturbation widths $\Delta{L}/L_{\rm n}$ (cf.~Eq.~\eqref{eq:skew} and the discussion around it) that generated the unsteady pulses are reported in the caption of Fig.~\ref{f:vmax}. In addition, in the manuscript it is stated that the time it takes each pulse from formation to hitting the line is of order ${\cal T}(\tau_d)$. The actual times are reported in the caption of Fig.~\ref{f:vmax}.

\subsection{Full fields comparison}

\begin{figure}[htp]
\includegraphics[width=0.9\columnwidth]{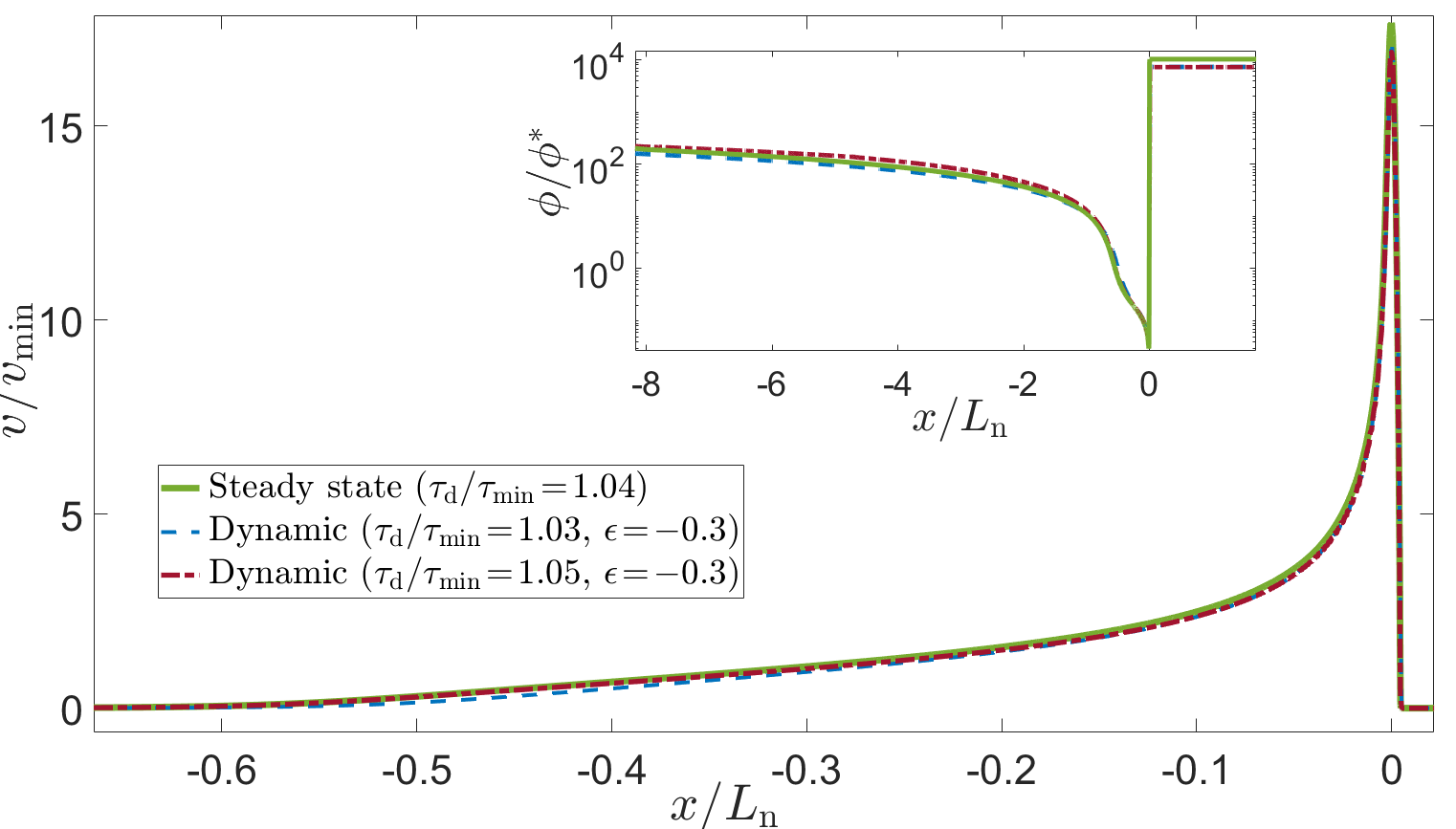}
\caption{\label{f:ss_dyn} The steady-state slip velocity field $v_{_{\rm ss}}\!(\xi)$ (with $\xi\!\to\!x)$ for $\tau_{\rm d}/\tau_{\rm min}\!=\!1.04$ (solid green line). Superimposed are instantaneous snapshots of $v(x,t)$ generated with $\tau_{\rm d}/\tau_{\rm min}\!=\!1.03,\,\epsilon\!=\!-0.3$ (dashed blue line, a decaying pulse) and $\tau_{\rm d}/\tau_{\rm min}\!=\!1.05,\,\epsilon\!=\!-0.3$ (dash-dotted brown line, a growing pulse) at times $t$ for which each unsteady pulse trajectory coincides with the steady-state point $(L^{\mbox{\tiny{(0)}}}(\tau_{\rm d}\!=\!1.04\tau_{\rm min}),\, c^{\mbox{\tiny{(0)}}}_{\rm p}(\tau_{\rm d}\!=\!1.04\tau_{\rm min}))$. (inset) The same for $\phi(x,t)$, on a semi-logarithmic scale (the $x$-axis is truncated at large negative values). The results reveal great quantitative agreement between the fields of the unsteady pulses (both growing and decaying) and their steady-state counterpart.}	
\end{figure}

The results presented in Figs.~3-4 in the manuscript and in Fig.~\ref{f:vmax} here suggest that the deep relation between unsteady slip pulses and their steady-state counterparts is valid not just at the level of discrete, coarse-grained variables that characterize the dynamical fields in the problem, but rather to the fields $v(x,t)$ and $\phi(x,t)$ (which of course include infinitely many degrees of freedom) themselves. That is, we expect instantaneous snapshots of $v(x,t)$ and $\phi(x,t)$ of unsteady pulses (whether growing or decaying) to be very similar to their steady-state counterparts $v_{_{\rm ss}}\!(\xi)$ and $\phi_{_{\rm ss}}\!(\xi)$ for some $\tau_{\rm d}$, once the latter is selected such that it is the $\tau_{\rm d}$ that corresponds to the point on the steady-state $c^{\mbox{\tiny{(0)}}}_{\rm p}[L^{\mbox{\tiny{(0)}}}]$ line at which the unsteady pulse trajectory resides at time $t$. This expectation is verified in Fig.~\ref{f:ss_dyn} for two unsteady pulses, one growing and one decaying, which happened to reside close to the steady-state $c^{\mbox{\tiny{(0)}}}_{\rm p}[L^{\mbox{\tiny{(0)}}}]$ line at some time during their dynamics.

\end{document}